\documentclass[sigconf, authorversion]{acmart}

\usepackage{appendix}
\usepackage{lipsum}
\usepackage{subcaption}
\usepackage{macros}
\usepackage{csquotes}
\usepackage{soul}

\def\change#1{#1}

\AtBeginDocument{%
  \providecommand\BibTeX{{%
    \normalfont B\kern-0.5em{\scshape i\kern-0.25em b}\kern-0.8em\TeX}}}

\copyrightyear{2023}
\acmYear{2023}
\setcopyright{rightsretained}
\acmConference[UIST '23]{The 36th Annual ACM Symposium on User Interface Software and Technology}{October 29-November 1, 2023}{San Francisco, CA, USA}
\acmBooktitle{The 36th Annual ACM Symposium on User Interface Software and Technology (UIST '23), October 29-November 1, 2023, San Francisco, CA, USA}
\acmDOI{10.1145/3586183.3606756}
\acmISBN{979-8-4007-0132-0/23/10}

\begin{document}

\title[Sensecape: Enabling Multilevel Exploration and Sensemaking with Large Language Models]{Sensecape: Enabling Multilevel Exploration and Sensemaking with Large Language Models}

\author{Sangho Suh}
\orcid{0000-0003-4617-5116}
\affiliation{%
  \institution{University of California, San Diego}
  \city{La Jolla}
  \country{USA}
}
\email{sanghosuh@ucsd.edu}

\author{Bryan Min}
\orcid{0009-0003-0657-4398}
\affiliation{%
  \institution{University of California, San Diego}
  \city{La Jolla}
  \country{USA}
}
\email{bdmin@ucsd.edu}

\author{Srishti Palani}
\orcid{0000-0003-1805-7307}
\affiliation{%
  \institution{University of California, San Diego}
  \city{La Jolla}
  \country{USA}
}
\email{srpalani@ucsd.edu}

\author{Haijun Xia}
\orcid{0000-0002-9425-0881}
\affiliation{%
  \institution{University of California, San Diego}
  \city{La Jolla}
  \country{USA}
}
\email{haijunxia@ucsd.edu}

\renewcommand{\shortauthors}{Sangho Suh, Bryan Min, Srishti Palani, Haijun Xia}

\begin{teaserfigure}
  \includegraphics[trim=0cm 0cm 0cm 0cm, clip=true, width=\textwidth]{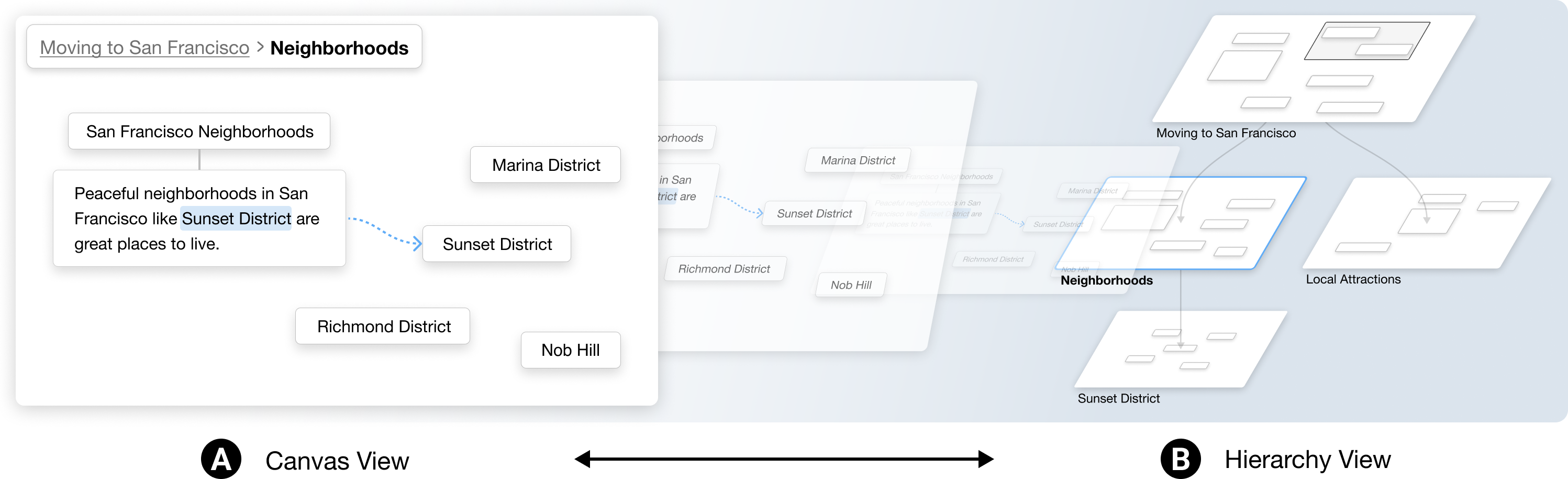}
  \caption{Sensecape allows users to seamlessly switch between the (A) canvas view and (B) hierarchy view for multilevel exploration and sensemaking. The hierarchy view embodies and externalizes the concept of levels of abstraction, providing a comprehensive overview of the information space and enabling users to navigate across different levels of abstraction for multilevel exploration and organize the collection of information at various levels of abstraction for sensemaking.}
  \Description{Teaser figure}
  \label{fig:teaser}
\end{teaserfigure}

\begin{abstract}
People are increasingly turning to large language models (LLMs) for complex information tasks like academic research or planning a move to another city. However, while they often require working in a nonlinear manner --- e.g., to arrange information spatially to organize and make sense of it, current interfaces for interacting with LLMs are generally linear to support conversational interaction. To address this limitation and explore how we can support LLM-powered exploration and sensemaking, we developed Sensecape, an interactive system designed to support complex information tasks with an LLM by enabling users to (1) manage the complexity of information through multilevel abstraction and (2) seamlessly switch between foraging and sensemaking. Our within-subject user study reveals that Sensecape empowers users to explore more topics and structure their knowledge hierarchically, thanks to the externalization of levels of abstraction. We contribute implications for LLM-based workflows and interfaces for information tasks.
\end{abstract}



\maketitle

\begin{figure*}[tb!]
	\centering
	\includegraphics[width=\textwidth]{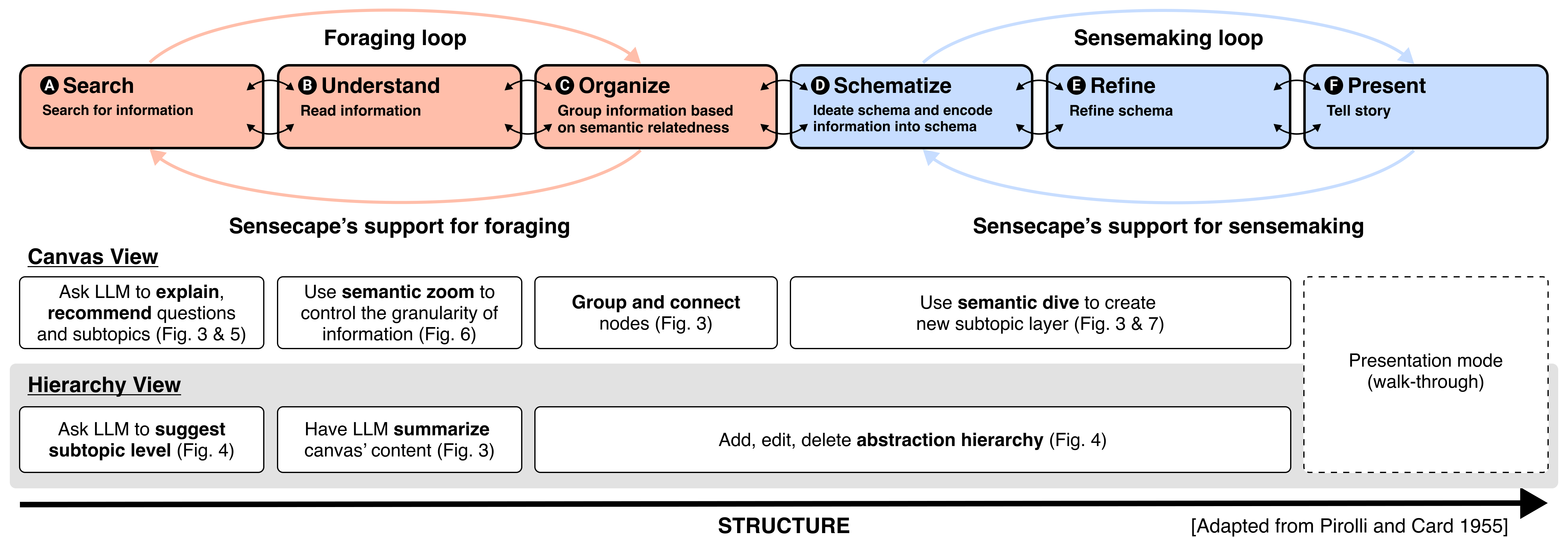}
	\caption{Information-seeking activities (top) alternate between foraging (exploration) and sensemaking loops~\cite{rachatasumrit2021forsense} and require encoding information into a structure as one moves from foraging to the sensemaking stage. How Sensecape supports each step (e.g., Search) in canvas and hierarchy views are shown (bottom), with the dotted line indicating the support not yet implemented.}
	\Description[short description]{long description.}
	\label{fig:framework}
\end{figure*}

\section{Introduction}
\label{section:introduction}

Large language models (LLMs) are revolutionizing the way we engage in information-related tasks. Millions of users are now turning to LLMs (e.g., ChatGPT) to find explanations, write essays, and summarize content, among many tasks, thanks to their ability to instantly generate high-quality responses to flexible natural language queries. 
The application of perhaps the broadest impact might be changing the way people obtain and make sense of information.
Instead of searching and browsing using search engines, people can converse with LLMs to acquire the desired information.

Although conversation is the most natural communication format, its inherent linear structure poses significant limitations for complex information tasks~\cite{langer1991discursive}. Linear conversational interfaces can be sufficient in supporting short question-answering tasks, such as responding `{\small\fontfamily{lmss}\selectfont 1 to 2 hours}' to the prompt \textit{``how long does it take to drive to San Francisco from San Jose?''} But, they are ineffective in assisting in complex information tasks that require gathering, organizing, and synthesizing information in a nonlinear manner. Take a trip planning task as an example. When presented with a list of places from ChatGPT in response to the prompt, \textit{``I am considering moving to San Francisco. What neighborhoods should I visit?''}, a user may want to get more recommendations, learn about a few locations in detail, revisit a previous list of recommendations for San Jose, or compare places between the two cities. However, to perform these tasks with the linear organization of the conversation history, the user has to navigate back and forth, which can lead to users quickly losing track of the overall information activity ~\cite{popolov2000conversation}. 


The challenges mentioned above arise from the fundamental mismatch between the sequential nature of a linear conversation and the highly flexible workflows and organizational approaches one utilizes when engaging in complex information tasks. Russell et al.~\cite{russell1993cost} suggests that \textit{``to answer task-specific questions, [we] search for a representation and encode data in that representation.''} As further highlighted in Fig.~\ref{fig:framework}, information work requires alternating between two main loops --- foraging loop and sensemaking loop. After exploring the information, people must find a good representation to encode it. Without support for flexible organization of the gathered information, making sense of the information can be challenging. In light of recent advances in LLMs and their growing presence in information-related tasks, these challenges can be exacerbated as LLMs become faster at responding to even complex queries and instantly generate large amounts of information. Therefore, the goal of this research is to reconcile this mismatch and enable a fluid exploration and sensemaking workflow with LLMs by exploring how they should be integrated with the diverse structures often employed in information tasks.


We demonstrate a way to address these challenges with Sensecape,\footnote{\url{https://sensecape.github.io}} an interactive system that enables users to engage in exploratory search tasks with LLMs by enabling multilevel exploration and sensemaking. Specifically, as shown in Fig.~\ref{fig:teaser}, Sensecape allows users to switch between the canvas and hierarchy views to help them explore and reason at different levels of abstraction by externalizing the abstraction hierarchy and enabling flexible navigation across these levels.

In summary, our work contributes to the development of new interfaces for LLMs that enable users to engage in information-seeking tasks in a more structured and systematic manner, providing a more comprehensive representation of the information space for sensemaking through multilevel abstraction. We contribute:

\begin{itemize}
    \item Sensecape, an interactive system that leverages the flexibility of a nonlinear interface suitable for exploratory tasks and the ability to flexibly navigate between the levels of abstraction in the information space;
    \item Externalization of multilevel abstraction for a more comprehensive and effective exploration of the information space for sensemaking;
    \item A user study demonstrating that enabling seamless exploration of semantic levels motivates and enables users to explore information space in an efficient and comprehensive manner.
\end{itemize}

\section{Related Work}
\label{section:related_work}

This research is inspired by user studies of information exploration and sensemaking and builds on the tools designed to support them.  

\subsection{Information Exploration and Sensemaking}
Complex information activities are often the interleave of two major tasks: exploratory search and sensemaking. When researching a complex and multifaceted subject matter, individuals tend to issue a series of iterative queries, scan and evaluate various information sources, and synthesize information from disparate sources to gain a better understanding of the topic, before incorporating it into their personal or professional knowledge bases~\cite{marchionini2006exploratory, white2009exploratory}. This has been referred to as \textit{exploratory search} in prior literature~\cite{white2009exploratory}. This open-ended exploration and discovery of information is the opposite of searching for a specific answer or piece of information. It is commonly used in situations when the searcher has limited prior knowledge or experience with the subject matter and needs a deeper understanding before proceeding with a more focused search.

To make sense of and work in complex, multifaceted information spaces, people take notes, curate relevant information \change{(Fig.~\ref{fig:framework}C)}, and create representations\change{/schema} (\change{Fig.~\ref{fig:framework}D;} e.g., tables,  graphs, concept maps, etc.) \change{to encode that information into them}. This process of encoding information into external representations to answer complex questions is known as \textit{sensemaking}~\cite{pirolli2005sensemaking}. This can free their mind from having to recall everything, help mentally process and synthesize all the information, and better reflect on the myriad ways the multiple facets are interconnected at different levels of abstraction~\cite{crescenzi2019towards, crescenzi2021supporting}.  

The complex and uncertain nature of this work makes it a nonlinear and dynamic process. Specifically, it involves switching back and forth between deduction and induction~\cite{budd2004relevance}, balancing divergent and convergent thinking~\cite{ford1999information}, and reflection along different levels of abstraction~\cite{foster2003serendipity}. Externalizing and reflecting on how the multiple levels and facets of information are interconnected or interdependent can be cognitively overwhelming and time-consuming. Furthermore, since the tools in which users take notes and work with this information are separate from the tools we use to explore information, it can be distracting to switch attention back and forth between the search and sensemaking tools~\cite{capra2010tools, meyer2014software}. For example, an academic literature review can take anywhere from a few hours to several months and include finding, reading, and making sense of anywhere between 30-50 sources, depending on the complexity of the topic and the depth of the review~\cite{knopf2006doing, denney2013write}.

\subsection{Tools for Information Exploration and Sensemaking}

\change{Recent work has started to integrate exploratory search and sensemaking. For example, InkSeine~\cite{hinckley2007inkseine}, Google Docs, and Microsoft Word allow people to issue words and annotations in their notes as queries. Recently, Microsoft and Google announced tools like CoPilot that allow users to ask questions in a chat sidebar next to the work application~\cite{bubeck2023sparks}. However, these methods still rely on users to identify and articulate their information needs as queries and do not guide the searcher to further explore their knowledge gaps or how to integrate the relevant information into their current knowledge. Research systems like CoNotate build on this and offer query suggestions based on the analysis of the searcher’s notes and previous searches~\cite{palani2021conotate}. Similarly, ForSense suggests parts of web pages to be clipped and clustered based on information the user has previously gathered~\cite{rachatasumrit2021forsense}. InterWeave builds on these systems by leveraging the content of the user’s sensemaking and embedding contextual suggestions into the user's evolving schema and knowledge structures~\cite{palani2022interweave}. Sensecape extends these work by further interweaving exploration and making sense of information across multiple levels of abstraction in an information space.}

\change{Over the past decades, Information Retrieval, NLP, and HCI researchers explored ways to support comprehensive exploration of information spaces~\cite{white2009exploratory}. To help find relevant information, prior research explored suggesting relevant topics, terms, or questions (e.g., `Related Searches' and `People also ask for' suggestions in search engines~\cite{rosset2020leading, baeza2004query}). To help navigate relevant information returned by search engines, faceted search interfaces~\cite{hearst2006clustering} employed categorization or clustering of search suggestions and results. To support the collection of relevant information, previous work explored highlighting and note-taking~\cite{roy2021note}, capturing information using clipping or bookmarking~\cite{kuznetsov2022fuse}, clustering clipped information~\cite{cutting2017scatter, chang2019searchlens, rachatasumrit2021forsense} and re-finding information~\cite{morris2008searchbar, dumais2003stuff}. More recently, large language models (e.g., GPT-4) demonstrated the unique ability to synthesize and generate information from large amounts of training data~\cite{bubeck2023sparks}. They can instantly provide the desired information to users and help with more complex information goals. Sensecape leverages these advances in LLMs and an understanding of cognitive strategies to build an intelligent 3D digital whiteboard that can help information workers explore and make sense of any topic.}


\subsection{Visuo-Spatial Organization of Information}
When working with complex information, people tend to organize information in their mind or externally, on notes or in their physical space (e.g., sticky notes, piles of paper)~\cite{kirsh2001context, henderson2009personal, malone1983people, jones1986spatial}. This visuo-spatial organization can help not only reduce the cognitive overload of storing everything in memory, share memory and mental context across information work sessions and collaborators~\cite{crescenzi2019towards, jones1986spatial, morris2013collaborative}, but also mentally manipulate complex information, and solve problems. It enables us to abstract complex and rich information and represent it in more manageable forms of representation so that we can apply spatial reasoning, such as rotation and transformation~\cite{shepard1971mental, huttenlocher1973mental}, highlight spatial relationships between information~\cite{tversky2005visuospatial}, identify patterns and symmetries, and solve problems creatively~\cite{tversky2005visuospatial, owens2014visuospatial}. 

Organizing information in a 3D space --- as in, for example, the hierarchy view (Fig.~\ref{fig:teaser}) --- additionally enables us to think across different levels of abstraction and encode a more accurate and intuitive relationship between different pieces of information~\cite{tversky2005visuospatial, jones1986spatial}. Especially when it comes to complex, multifaceted information or information related to physical or real-world objects. SemNet~\cite{fairchild2013semnet}, the Information Visualizer project at Xerox PARC~\cite{card1991information}, Workscape~\cite{ballay1994designing}, and Data Mountain~\cite{robertson1998data} were early systems that introduced the 3D spatial layout of documents. The Web Forager~\cite{card1996webbook} built on this work introduced a 3D spatial layout for web pages. The automatic spatial layouts of information in these systems leverage the user’s ability to recognize and understand spatial relationships (both in 2D and 3D). The 3D interface makes it possible to display more information without incurring an additional cognitive load~\cite{robertson1998data, tversky2005visuospatial}. 

\change{If not designed carefully, however, 3D interfaces can create problems than provide benefits~\cite{cockburn2002evaluating}. For example, prior work found that while users are enthusiastic about using 3D interfaces, they can find them difficult to use until they gain enough experience. Specifically, when 3D interfaces have \textit{depth} --- with objects varying in size depending on their distance from the screen --- and allow objects to obscure one another (\textit{occlusion}), they can make the interface \textit{cluttered} and challenging to use --- e.g., when users are trying to locate and select objects~\cite{cockburn2000evaluation}. To avoid such issues, we designed the hierarchy view to be pseudo-3D, with minimal use of \textit{depth} and no canvas layers \textit{occluding} one another.}

\change{Prior studies comparing 2D and 3D interfaces for working with information also offer a number of insights on when to use 2D and 3D interfaces. They showed that user performance and experience for 2D and 3D interfaces largely depend on the combination of task, user, and interface and that these should be carefully considered~\cite{sebrechts1999visualization, tory2007spatialization}. Overall, Sensecape builds on this rich literature of cognitive theories and prior systems to support the nonlinear, iterative, and dynamic nature of complex information work by externalizing information at different levels of granularity and hierarchy, and highlighting inter-dependencies across these levels.}

\subsection{Managing Complexity of Information Space}
HCI research developed and studied the effects of various interaction techniques for managing the complexity of information space: 

\begin{itemize}
    \item \textit{Semantic zoom} allows users to zoom in and out of a visual representation of the information space, enabling them to focus on specific details or get an overview of the entire space. Zooming can enhance users' sense of control and help them better understand the spatial relationships between objects~\cite{bederson1994pad++, ware1990exploration}.
    \item \textit{Filtering} allows users to selectively display information based on certain criteria or parameters, making it easier to focus on relevant information. It has been found to be an effective technique for reducing information overload and improving task performance~\cite{hearst1992automatic}. Similarly, \textit{clustering and categorization} group similar items to reduce the complexity of a dataset. For example, a user might group similar products on an e-commerce website or cluster similar documents in a search engine.
    \item \textit{Navigation} allows users to move around the information space, enabling them to explore different parts of the space and access relevant information. Effective navigation can improve users' comprehension of complex information spaces and reduce their cognitive load~\cite{benyon2001new, rivlin1994navigating}.
    \item \textit{Linking and brushing} technique involves linking multiple views of the same data, allowing users to see how changes in one view affect others. For example, a user might brush a region of a scatterplot to highlight the corresponding data points in other views. Research has shown that linking and brushing can improve users' ability to find patterns and relationships in data~\cite{becker1988use}.
    \item \textit{Flexible representations} combine some of the above interaction techniques to support information exploration and processing. For example, WritLarge~\cite{xia2017writlarge} integrates pinch-to-zoom and selection in a single gesture for fluidly selecting and acting on content; and addresses the combined issues of navigating, selecting, and manipulating content by allowing the transformation of information across semantic, structural, and temporal axes.
\end{itemize}

Sensecape implements these interaction techniques to support exploration, reasoning around and management of complex information spaces.

\section{Sensecape}
\label{section:system}

\begin{figure*}[htb!]
	\centering
	\includegraphics[width=\textwidth]{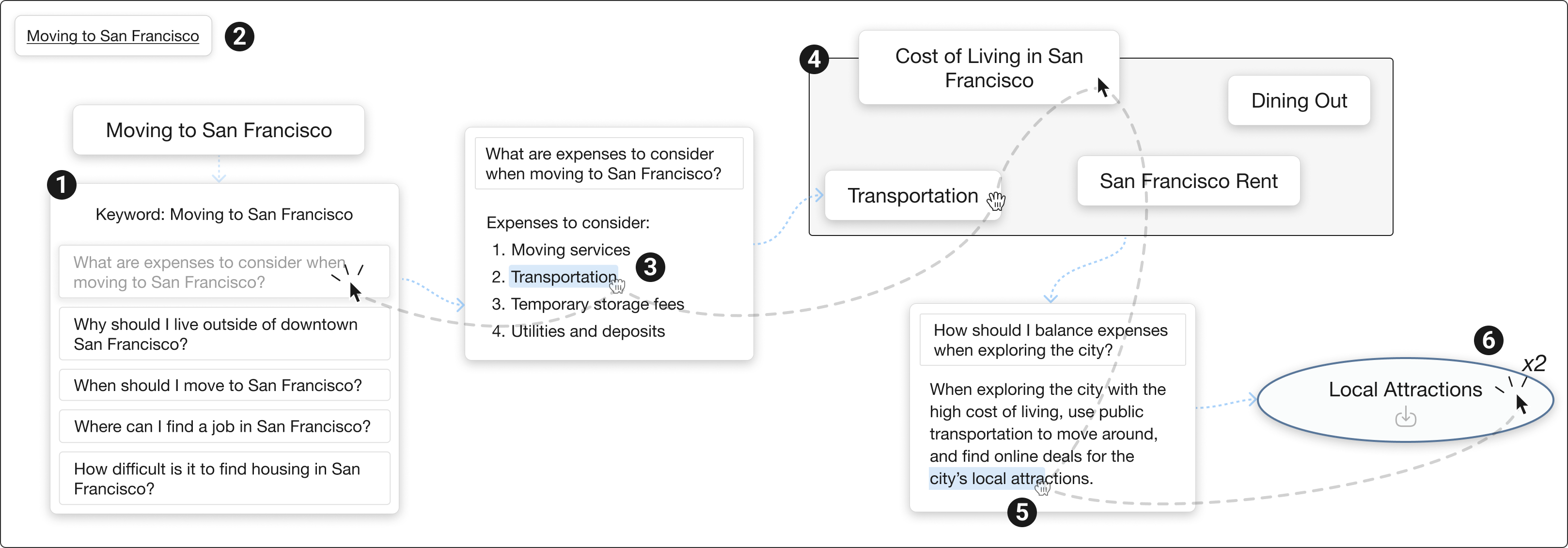}
 	\caption{An example workflow on canvas view. A user asks Sensecape to (1) generate a list of questions by selecting the node `{\small\fontfamily{lmss}\selectfont Moving to San Francisco}' and the \textsc{Questions} button in expand bar (Fig.~\ref{fig:expand-bar}); Sensecape (2) updates the canvas topic to `{\small\fontfamily{lmss}\selectfont Moving to San Francisco}' based on the new information on the canvas; the user explores `{\small\fontfamily{lmss}\selectfont Expenses to consider}' when moving to San Francisco; the user (3) highlights `{\small\fontfamily{lmss}\selectfont Transportation}' to create a node and (4) groups it with other relevant topics (e.g., `{\small\fontfamily{lmss}\selectfont San Francisco Rent}', `{\small\fontfamily{lmss}\selectfont Dining Out}'), under their high-level topic `{\small\fontfamily{lmss}\selectfont Cost of Living in San Francisco}'. As LLM explains how to balance expenses when exploring the city, the word `{\small\fontfamily{lmss}\selectfont local attractions}' catches the user's eye. The user (5) drags the highlighted text out from the node to create its node and (6) double clicks it to \textit{dive} into it (\textit{Semantic Dive} in Section~\ref{section:user-interface}) to explore the topic in a separate canvas.}
	\Description[short description]{long description.}
	\label{fig:canvas-view}
\end{figure*}

To give a clear picture of the motivations underlying the design and features in Sensecape, we first present a motivating scenario, elucidating the \ul{challenges} (\textbf{C}s) that users encounter when performing complex information tasks with a conversational interface.

\subsection{Motivating Scenario}
\label{section:motivating_scenario}

Paul is a student considering moving to San Francisco after graduating. Since he has never lived there and does not know much about the city, he plans to do some research and visit in a few weeks. He decides to use ChatGPT for his research and trip planning. 

\noindent\textbf{\textsc{Exploration.}} Since he does not know much about the city, he \ul{is unsure where to begin and what to ask} (\textbf{C1}). After thinking, he decides to directly ask, ``\textit{I plan to move to San Francisco. What should I look for?}'' ChatGPT lists several things to consider: `{\small\fontfamily{lmss}\selectfont Location}', `{\small\fontfamily{lmss}\selectfont Cost of living}', `{\small\fontfamily{lmss}\selectfont Climate}', `{\small\fontfamily{lmss}\selectfont Culture}', `{\small\fontfamily{lmss}\selectfont Commute}', `{\small\fontfamily{lmss}\selectfont Housing options}', and `{\small\fontfamily{lmss}\selectfont Activities and entertainment}'. He finds all these topics relevant and worth investigating. However, since he can research only one topic at a time, he decides to explore `{\small\fontfamily{lmss}\selectfont Location}' which, in full, reads: `{\small\fontfamily{lmss}\selectfont Location: Some neighborhoods are more walkable than others. Consider the proximity of your potential new home to public transportation, grocery stores, restaurants, and other amenities you may need}.' This reminds him that he wants to live in a quiet neighborhood. So he asks, ``\textit{What are some quiet neighborhoods?}'' ChatGPT returns several neighborhoods: `{\small\fontfamily{lmss}\selectfont Forest Hill}', `{\small\fontfamily{lmss}\selectfont Outer Sunset}', `{\small\fontfamily{lmss}\selectfont Sea Cliff}', and so on. He asks a few follow-up questions about `{\small\fontfamily{lmss}\selectfont Sea Cliff}' --- realizing soon, however, that that \ul{the list of underexplored neighborhoods} (e.g., `{\small\fontfamily{lmss}\selectfont Forest Hill}', `{\small\fontfamily{lmss}\selectfont Outer Sunset}') \ul{is no longer visible due to the linear nature of the conversational interface} (\textbf{C2}). Although he knows that he still needs to check other neighborhoods, he decides to instead explore another subtopic (e.g., `{\small\fontfamily{lmss}\selectfont Cost of living}') because he does not want to exert effort to look for the list.

\noindent\textbf{\textsc{Sensemaking.}} After exploring the `{\small\fontfamily{lmss}\selectfont cost of living}' using ChatGPT, he realizes that he should document and synthesize the explored topics and gathered information. He also realizes that ChatGPT's responses at different points of the conversation are relevant and useful for different topics (e.g., cost of living for residents in Sea Cliff). He wants to \ul{group these related information but cannot do this} (\textbf{C3}) in the ChatGPT environment. He decides to use Miro, a digital whiteboard tool. He copies and pastes generated responses from the ChatGPT interface to the canvas. He zooms out to arrange them into groups, but \ul{too much text overwhelms him, making it difficult to identify key ideas in each response} (\textbf{C4}). He breaks down responses into several nodes containing just keywords (e.g., `{\small\fontfamily{lmss}\selectfont Location}', `{\small\fontfamily{lmss}\selectfont Cost of living}'), relevant information in summary (e.g., `{\small\fontfamily{lmss}\selectfont Consider the proximity to public transportation, grocery stores, and restaurants}'), and questions he wants answered (e.g., ``\textit{Which neighborhood is the best for a young single adult?}'').

He zooms out again, checks keywords or summaries of the text, identifies the relevance of each node to another, and places them together. Moreover, he notices that the collected information can form a hierarchy, with the main topic `{\small\fontfamily{lmss}\selectfont Moving to San Francisco}' at the top of the hierarchy and its subtopics from the first interchange (i.e., `{\small\fontfamily{lmss}\selectfont Location}', `{\small\fontfamily{lmss}\selectfont Cost of living}', `{\small\fontfamily{lmss}\selectfont Climate}', and others) in the sub-level. He moves nodes accordingly to form a hierarchical layout. After returning to ChatGPT, he becomes curious about nearby cities such as `{\small\fontfamily{lmss}\selectfont San Jose}'. He decides to consider them as possible destinations as well. Back in Miro, since this is not a subtopic under `{\small\fontfamily{lmss}\selectfont San Francisco}' (now edited from `{\small\fontfamily{lmss}\selectfont Moving to San Francisco}'), he adds `{\small\fontfamily{lmss}\selectfont My Future Home}' above `{\small\fontfamily{lmss}\selectfont San Francisco}' and then places `{\small\fontfamily{lmss}\selectfont San Jose}' next to it at the same level. Unfortunately, the Miro \ul{workspace quickly gets cluttered as he expands his hierarchy with new subtopics and information} (\textbf{C5}). \ul{The context-switching between two systems and transferring the generated responses to the workspace also becomes time-consuming and laborious.} (\textbf{C6}).

In summary, the challenges with conducting complex information tasks with a conversational interface are as follows:

\noindent\textbf{C1. Slow Start:} Users with limited knowledge of the topic or with complex information goals that require iterative exploration of various facets face challenges in determining where to start and which questions to ask.

\noindent\textbf{C2. Hard to Revisit}: Information organized in a linear sequence makes it challenging for users to track and revisit previous topics.

\noindent\textbf{C3. Lack of Structure:} The inability to group and specify connections across information makes sensemaking difficult.

\noindent\textbf{C4. Information Overload:} The large amount of information generated by LLMs can pose cognitive overload.

\noindent\textbf{C5. Visual Clutter:} When exploring multiple topics, the canvas can quickly get cluttered, making it challenging to understand the relationship between topics at a high level.

\noindent\textbf{C6. Cost of Context-Switching:} The disconnect between information exploration and sensemaking forces users to frequently switch contexts, which results in inefficient workflows.

\subsection{User Interface \& Features}
\label{section:user-interface}

\begin{figure*}[htb!]
	\centering
	\includegraphics[width=\textwidth]{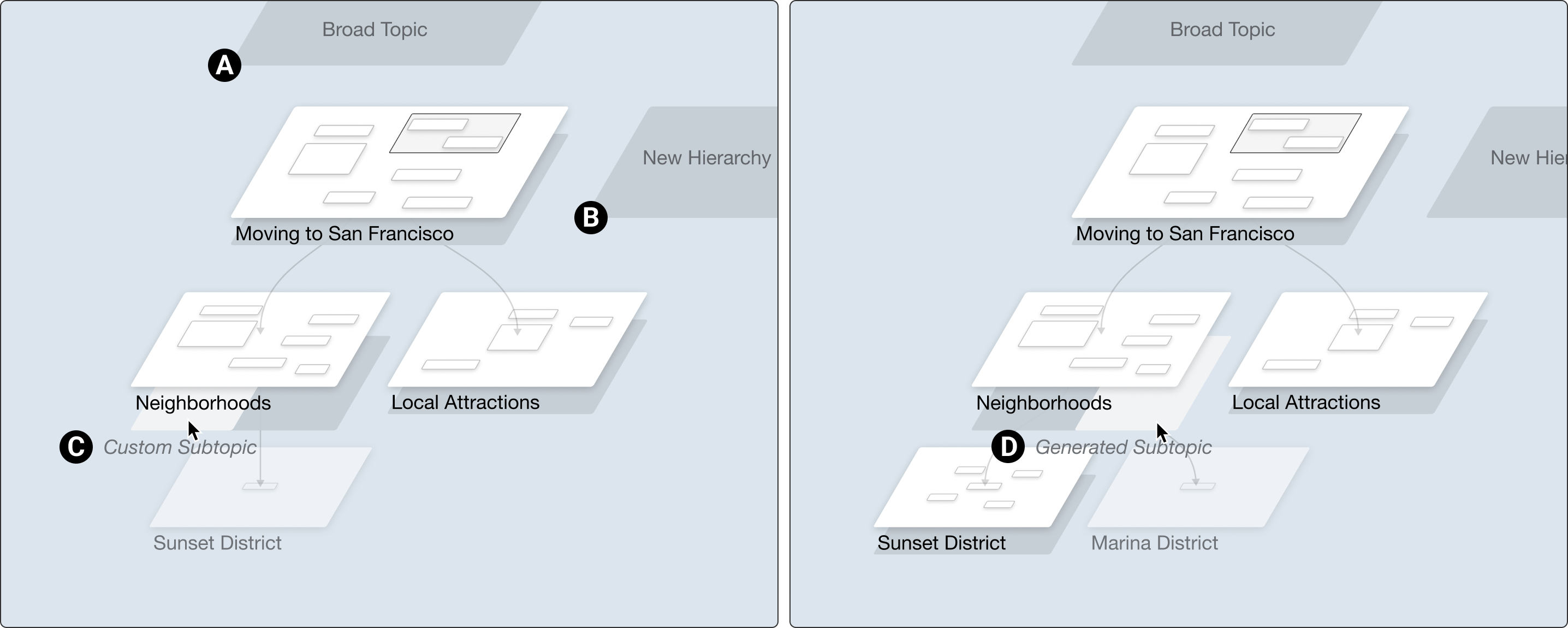}
	\caption{Hierarchy view: Users can add (A) a canvas above (e.g., `{\small\fontfamily{lmss}\selectfont Relocating to a new city}' above `{\small\fontfamily{lmss}\selectfont Moving to San Francisco}') or (B) another hierarchy on the side (e.g., `{\small\fontfamily{lmss}\selectfont Moving to San Jose}' next to `{\small\fontfamily{lmss}\selectfont Moving to San Francisco}'). To add a subtopic canvas, users can click (c) {\small\fontfamily{lmss}\selectfont\textsl{Custom Subtopic}} and specify the topic (e.g., `{\small\fontfamily{lmss}\selectfont Sunset District}') or (D) {\small\fontfamily{lmss}\selectfont \textsl{Generated Subtopic}} to have LLM suggest a subtopic (e.g., `{\small\fontfamily{lmss}\selectfont {\small\fontfamily{lmss}\selectfont Marina District}}').}
	\Description[short description]{long description.}
	\label{fig:hierarchy-view}
\end{figure*}

Sensecape consists of two main views: \textit{canvas} view (Fig.~\ref{fig:canvas-view}) and \textit{hierarchy} view (Fig.~\ref{fig:hierarchy-view}). The main difference is the semantic level at which users perform exploratory and sensemaking tasks, as shown in Fig.~\ref{fig:framework}. The canvas view allows users to search, gather, and organize information on any topic. The hierarchy view helps users reason at a higher level: users see each canvas, its topic, and where they are in relation to other topics in the 3-dimensional space and at which level of abstraction.  Below, we describe each view, its features, and how they help address the challenges (\textbf{C1}-\textbf{C6}).

\subsubsection{Canvas View [C2-4, C6]}
\label{section:canvas-view}
The canvas view is an infinite whiteboard where users can perform basic diagramming functionalities, such as adding, grouping, and connecting nodes with edges (\textbf{C3}). They can also search and organize the generated response (\textbf{C2}) directly on the canvas (\textbf{C6}). The canvas view follows the \textit{node-first approach}. The first step in any interaction begins with adding a node, which can be performed by double clicking anywhere on the canvas. After creating a node, the user can input text --- e.g., topic (e.g., `{\small\fontfamily{lmss}\selectfont San Francisco Culture}'), statement (e.g., `{\small\fontfamily{lmss}\selectfont San Francisco has a mild Mediterranean climate, but the weather can vary depending on the neighborhood.}'), or question (e.g., ``{\small\fontfamily{lmss}\selectfont Why should I live outside of downtown San Francisco?}''). As shown in Fig.~\ref{fig:canvas-view} (2), as new information is added to the canvas, Sensecape uses LLM to summarize the content on the canvas into a single topic and updates the topic on display at the top left corner (\textbf{C4}).

\begin{figure}[hbt!]
	\centering	\includegraphics[width=0.47\textwidth]{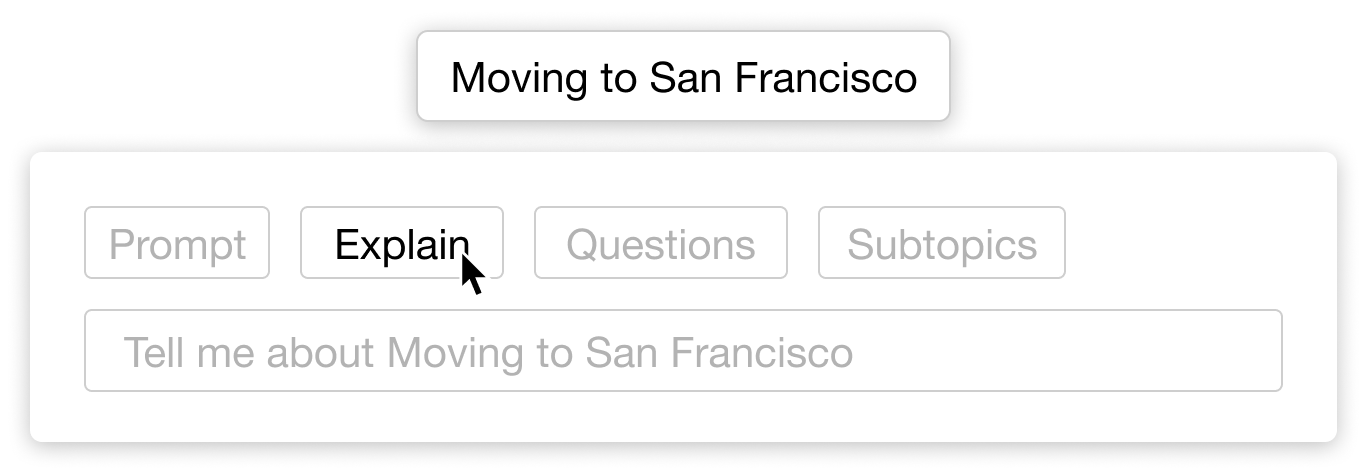}
	\caption{Expand bar: the cursor hovering over \textsc{Explain} previews the prompt in the input box as a placeholder. Clicking it sends this prompt to LLM and adds the response below.}
	\Description[short description]{long description.}
	\label{fig:expand-bar}
\end{figure}

\paragraph{Expand Bar [C1]}


Once a node is added, the user can click it and have the expand bar appear below the node. Expand bar offers several functionalities to help users' exploration and sensemaking. As shown in Fig.~\ref{fig:expand-bar}, expand bar allows users to use the node's text as a \textsc{Prompt} or as a basis for \textsc{Explain}, \textsc{Questions}, and \textsc{Subtopics} features. Concretely, if the node's text reads, ``\textit{What are the top San Francisco attractions?}'' and the user clicks \textsc{Prompt}, this question will be fed to an LLM. The LLM-generated response is then streamed to a node created below this node. \textsc{Questions} is designed to address the situation in which the user struggles to find out where to start and what questions to ask (\textbf{C1}). When the user clicks \textsc{Questions}, a node containing 25 questions --- generated by LLM and begin with \textit{What}, \textit{Why}, \textit{Where}, \textit{When}, and \textit{How} --- is added, as shown in Fig.~\ref{fig:canvas-view} (1). \textsc{Explain} helps expedite the exploration process by allowing users to quickly retrieve an explanation of a topic. Compared to \textsc{Prompt}, which uses the node's text directly as a prompt, \textsc{Explain} adds `{\small\fontfamily{lmss}\selectfont Tell me about}' in front of the prompt, as shown in Fig.~\ref{fig:expand-bar}. Finally, \textsc{Subtopics} generates subtopics around the node, facilitating the exploration process when the user is out of ideas on topics to explore (\textbf{C1}).

\paragraph{Text Extraction [C2, C3]}

To help users break down the generated response, users can highlight parts of the generated response and then either click or drag the highlighted text to the canvas to create a new node containing the highlighted text, as illustrated in Fig.~\ref{fig:canvas-view} (3). This allows users to extract a topic or information they find worth exploring further at a later time (\textbf{C2}) and position it at the desired location for organization (\textbf{C3}).

\begin{figure}[htb!]
    \begin{tabular}{lll}
        \subfloat[Semantic level: all]{\includegraphics[width = 0.42\textwidth]{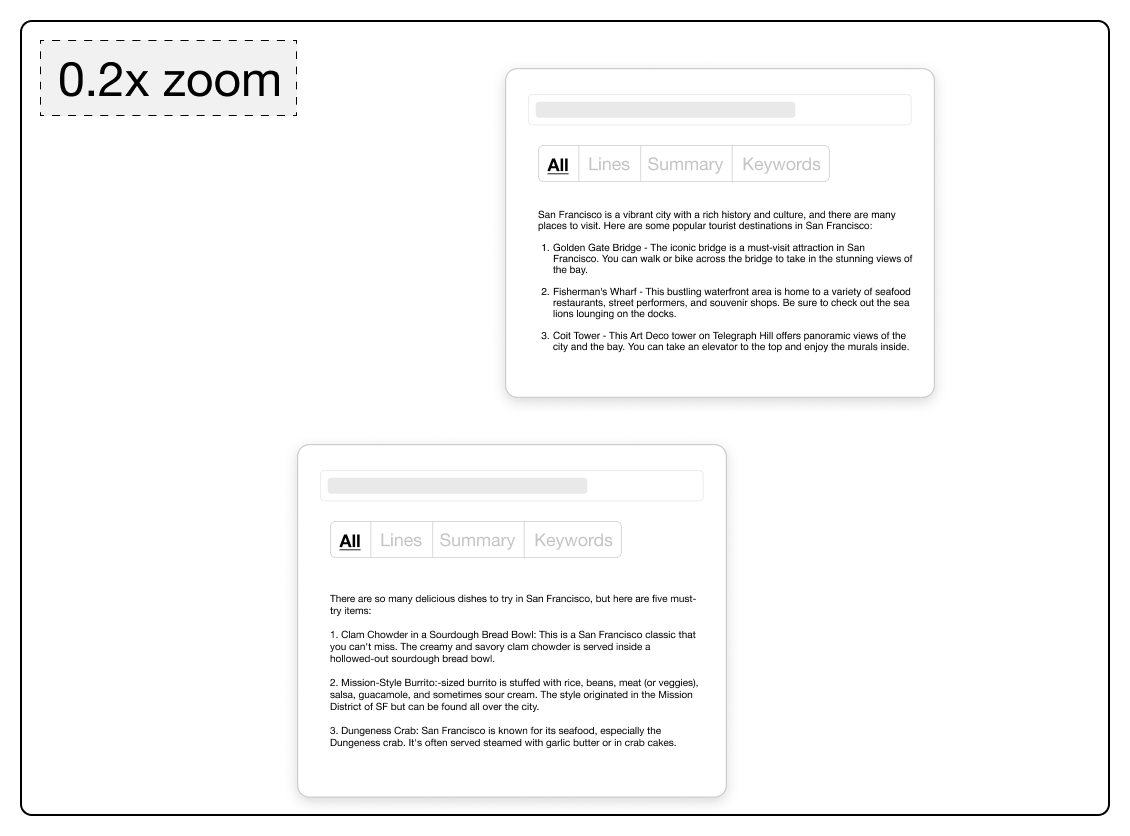}} \\
        \subfloat[Semantic level: keywords]{\includegraphics[width = 0.42\textwidth]{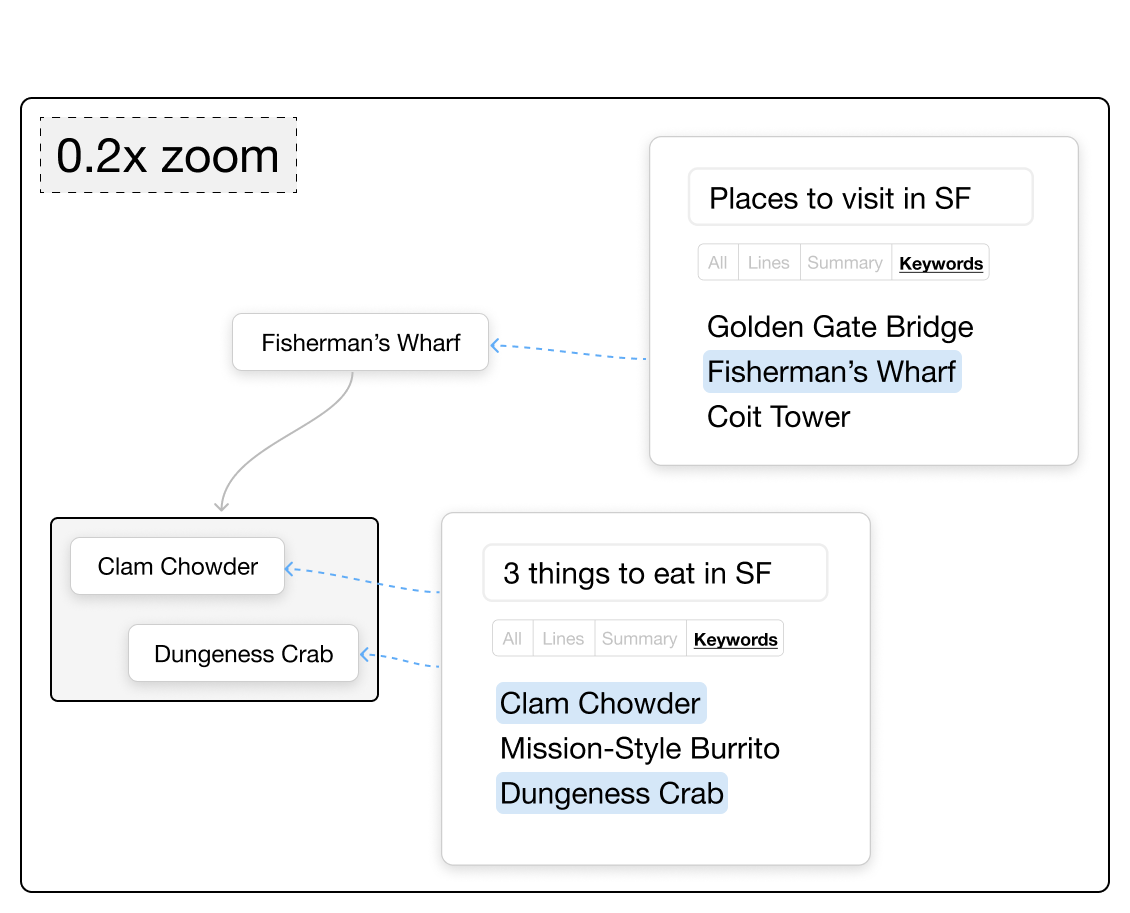}} 
    \end{tabular}
    \caption{Semantic zoom: Users can control the granularity of the information. When they zoom out (e.g., 0.2x) to see multiple nodes on canvas, the text in each node can be overwhelming and difficult to read, as shown in (a). In Sensecape, users can set the semantic level to, e.g., keywords level, to manage information overload and connect key topics within nodes, as shown in (b).}

    \label{fig:semantic-zoom}
\end{figure}

\paragraph{Semantic Zoom [C4]}
While LLMs enable us to retrieve answers instantly, the amount of text generated can be overwhelming. In Sensecape, users can use semantic zoom to manage this information overload (\textbf{C4}). By default, users will see the generated response as is. Then, as the user's zoom level changes (i.e., the user zooms in or out), the response will dynamically update, as shown in Fig.~\ref{fig:semantic-zoom}, to show keywords, for example, when users zoom out. This can be useful if many nodes are on the canvas and a user needs to easily identify the explored topics and connections among the nodes. If they want to manually set it to any semantic zoom level regardless of the zoom level, they also have the option. For example, they can select \textsc{Summary} or \textsc{Keywords} to see the summary version of the response or its keywords, respectively. This can be useful, for example, if they ask multiple questions to LLM and several responses are generated and are too long for the user to process. After Sensecape fetches the response from LLM, it feeds the response back to LLM and prompts it to return the response at different levels of detail --- (1) \textsc{Lines}: response with multiple paragraphs/lines summarized into summaries of each; (2) \textsc{Summary}: response with the entire response summarized into one paragraph; (3) \textsc{Keywords}: response with the entire response abridged into keywords. By default, the semantic zoom --- set at \textsc{Auto} --- changes the level of detail depending on the zoom level. For example, as the user zooms out, the semantic level progresses to the less detailed (e.g., \textsc{All} $\rightarrow$ \textsc{Summary} $\rightarrow$ \textsc{Keywords}) and reverts to the more detailed semantic level as the user zooms in to node (e.g., \textsc{Keywords} $\rightarrow$ \textsc{Summary} $\rightarrow$ \textsc{All}).

\begin{figure}[htb!]
	\centering	\includegraphics[width=0.45\textwidth]{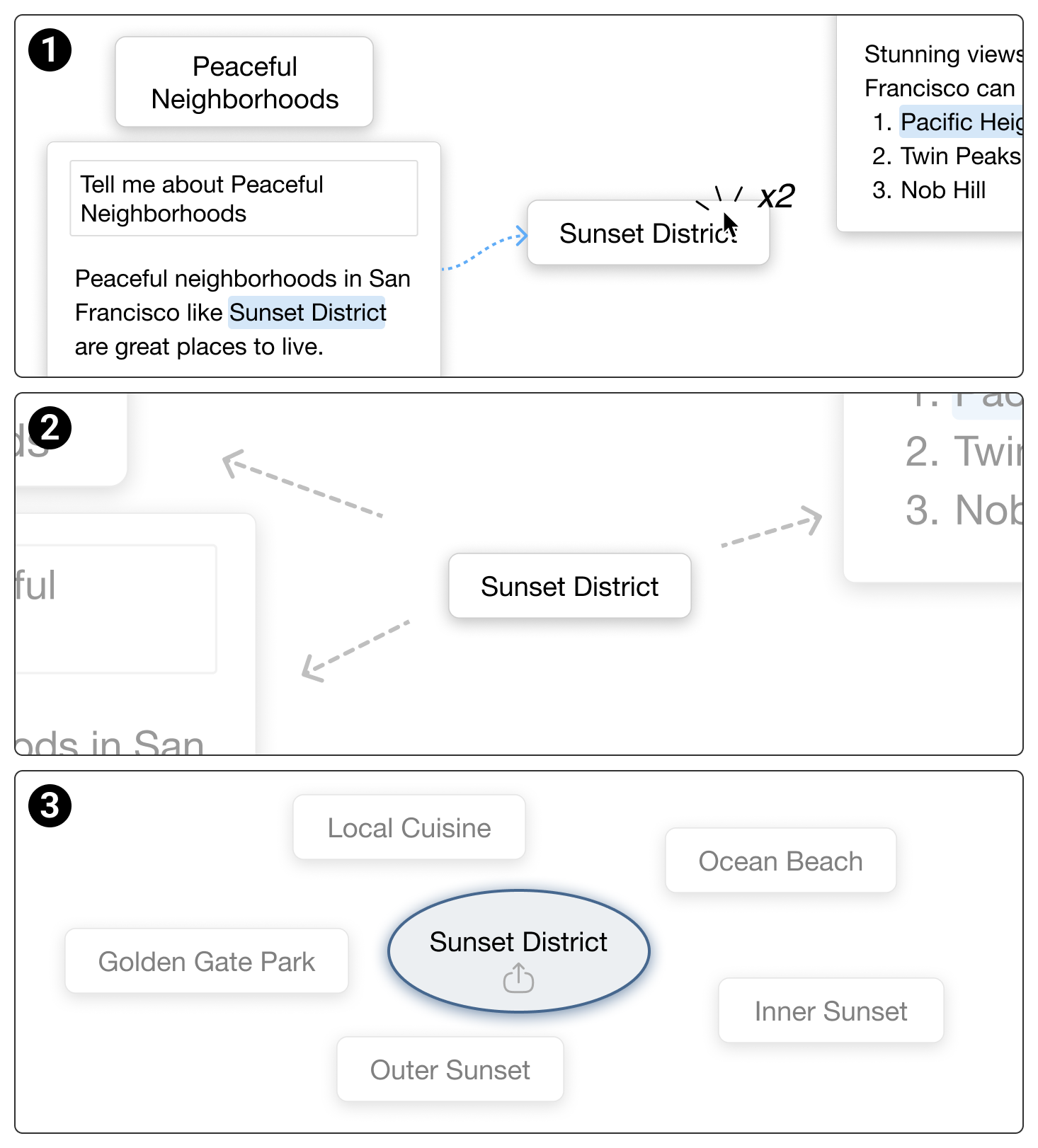}
	\Description[short description]{long description.}
	\caption{Semantic dive: Users can (1) dive deeper into the topic by double clicking on the node. (2) Surrounding elements are pushed away as the selected node is carried into a lower level canvas. In this new canvas, users are (3) automatically recommended subtopics of the topic doven into. Semantic dive also updates the hierarchy view, creating a lower level canvas matching the action in Figure \ref{fig:hierarchy-view} (C).}
	\label{fig:semantic-dive}
\end{figure}

\paragraph{Semantic Dive [C5]}
\label{feature:semantic-dive}

When users find the current working canvas too cluttered or a node they would like to explore further in a separate canvas, they can quickly \textit{dive} into that canvas by double clicking the node. When users double click the node, it transforms into an ellipse-shaped \textit{portal} node and takes them to a canvas layer below the canvas they were in. In other words, Sensecape creates an empty canvas layer for exploring that topic and pulls the user into that layer. Thus, in addition to being able to manually add the canvas layer in the hierarchy view, they can also do so in the canvas view by performing \textit{semantic dive} on any node. \change{If users want to transfer any nodes and user-defined schema across canvas layers when they perform semantic dive or move into a new layer, they can easily do so. This can be done by selecting any nodes on the canvas, copying (\textsc{ctrl + c}), and then pasting (\textsc{ctrl + v}) them onto a new canvas. As they are pasted, the structure is carried over, helping users with the organization and any re-structuring efforts.}

\subsubsection{Hierarchy View [C1-5]}

The hierarchy view (Fig.~\ref{fig:hierarchy-view}) offers a holistic view for users to identify where they are and what they are exploring in the context of the overall information space (\textbf{C2}). It provides an overview of the information space and allows users to reflect on the relation between the canvases and navigate to them. At the same time, it is a way for users to address the visual clutter on their canvases, as zooming out from the canvas view into the hierarchy view allows users to abstract and transform the content in each canvas into a more manageable form of representation \change{(\textbf{C4}, \textbf{C5}) --- not to mention it encourages users to structure the information hierarchically to assist their sensemaking (\textbf{C3}). As illustrated in Fig.~\ref{fig:hierarchy-view}, users can also ask LLMs to recommend a potential subtopic they can explore next (\textbf{C1}).}

\change{The hierarchy view is inspired by our pilot studies and prior work on the multi-window approach~\cite{henderson1986rooms}. Our pilot studies revealed that since LLM generates a large amount of text, it can quickly clutter the canvas space during its use. The multi-window approach showed that to address the information overload, we can assign information clusters into their own spaces. Thus, to mitigate the visual clutter issue (\textbf{C5}) and simultaneously support the schematization (Fig.~\ref{fig:framework}D) of hierarchical relationships between topics and their subtopics, we represented dedicated topic spaces in a hierarchy (\textbf{C3}).}

\paragraph{Adding \& Deleting Higher and Lower Level Canvas [C2-3, C5]} Users can construct their hierarchy by adding and deleting canvases. Users can add a higher-level topic canvas by clicking the \textsc{Broad Topic} button shown in Fig.~\ref{fig:hierarchy-view} (A), and users can add lower-level subtopic canvases by clicking the shadow underneath the canvases. On hover, the subtopic button reveals two options for users to add either a custom subtopic or a subtopic generated by LLM, as shown in Fig.~\ref{fig:hierarchy-view} (C) and (D) respectively. Users can remove canvases and entire canvas branches to prune their exploration space (\textbf{C5}). By allowing users to add and edit canvases in the hierarchy view, users can assign relations between canvases (\textbf{C2}) and simultaneously build their exploration and sensemaking space (\textbf{C3}).


\paragraph{Creating New Hierarchy [C3]} In addition to expanding upon the current hierarchy, users can form new hierarchies as shown in Fig.~\ref{fig:hierarchy-view} (B). This new hierarchy can be expanded with the same construction methods shown in the same figure. Users can also create connections between hierarchies by adding a higher-level topic canvas, as shown in Fig.~\ref{fig:hierarchy-view} (A). Once a higher-level topic canvas is added, Sensecape adds arrows from the new higher-level canvas to each hierarchy to form a single hierarchy (\textbf{C3}).

\subsection{Implementation Details}

Sensecape is a web application developed with React. The canvas was implemented using Reactflow, an open-source library for building diagramming applications. 

Sensecape's generative chat feature used OpenAI's `gpt-3.5-turbo' model, while all other LLM-based features used `gpt-4'. The `gpt-3.5-turbo' model was specifically used for chat responses as it generated content faster than `gpt-4'. All other LLM-based features prioritized accurate interpretation of our prompts (e.g., to generate the most relevant subtopics of a topic). Thus we used the `gpt-4' model. The prompts used are shown in Table~\ref{table:llm-prompts} in Appendix.

\begin{figure*}[htb!]
	\centering	\includegraphics[width=0.88\textwidth]{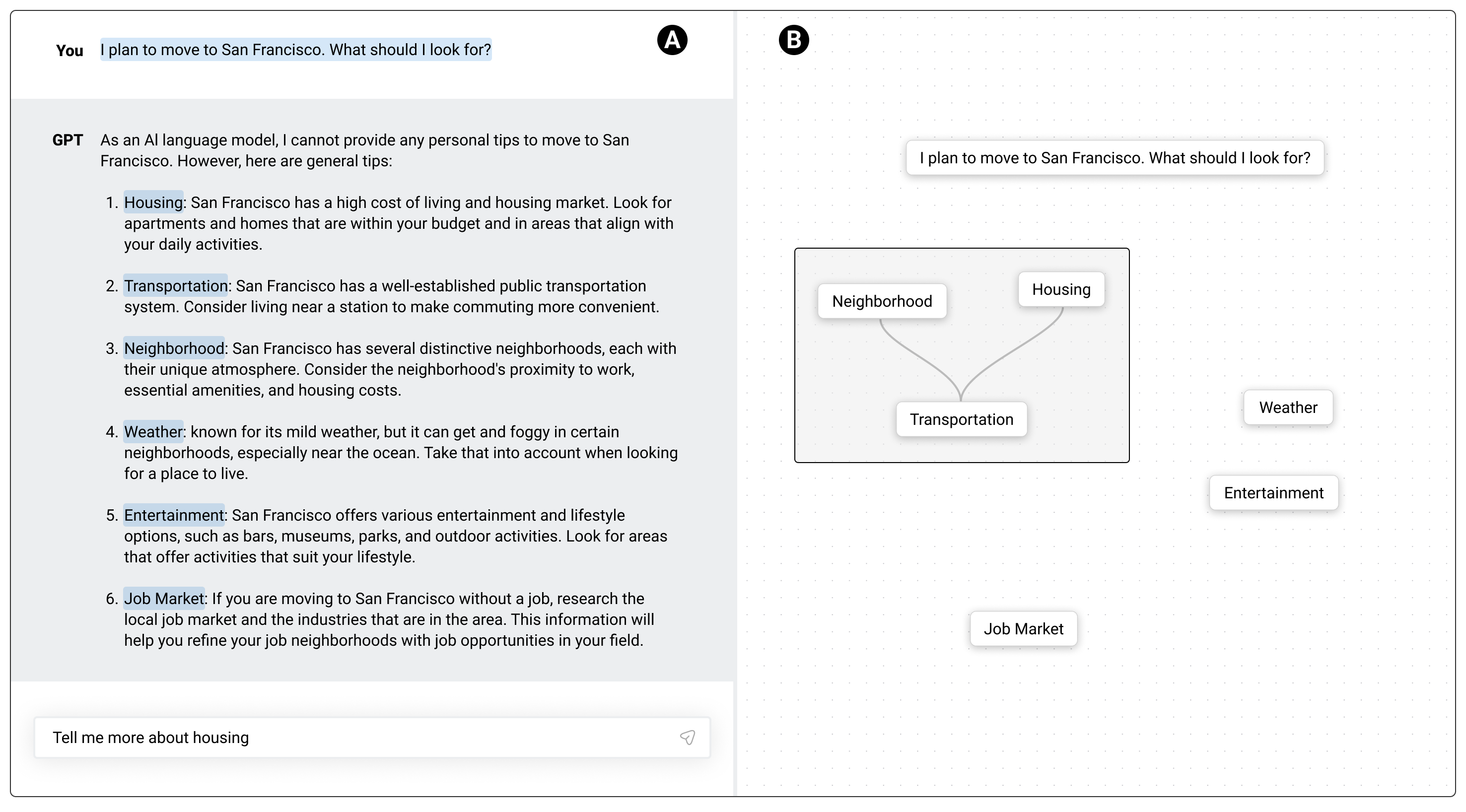}
	\caption{Baseline interface: (A) conversational interface resembling OpenAI's ChatGPT interface allowed participants to ask questions and issue prompts; (B) canvas view allowed participants to create, group, and connect nodes with edges. Participants could highlight any text in the conversational interface and click or drag the text to add them to the canvas.}
	\Description[short description]{long description.}
	\label{fig:baseline-interface}
\end{figure*}

\section{User Evaluation}
To evaluate whether Sensecape supports exploration and sensemaking, we conducted a within-subject study. Specifically, we aimed to answer the following questions:
\begin{itemize}
    \item \textbf{RQ1.} \change{How does} Sensecape support exploration?
    \item \textbf{RQ2.} \change{How does} Sensecape support sensemaking?
    \item \textbf{RQ3.} What is the perceived utility of Sensecape's features?
    \item \textbf{RQ4.} How do people see Sensecape being useful in their everyday knowledge work?
\end{itemize}


\subsection{Conditions} To assess the usefulness of the features and interactions that we introduce, we set our Baseline interface to be an integrated environment with a conversational interface and canvas. As shown in Fig.~\ref{fig:baseline-interface}, the left half of the interface features a conversational interface similar to ChatGPT, while the right half provides a canvas with diagramming capabilities. Users could interact with the conversational interface in the same manner as they would with ChatGPT's interface. The canvas on the right side served as a note-taking area. They could easily add parts of the text generated by LLM in the conversational interface region by highlighting text (including their prompt), clicking the highlighted text, or dragging it out to add a node containing the highlighted text to the canvas on the right. 
Baseline lacked the hierarchy view and had only basic diagramming functionalities such as adding, grouping, and connecting nodes. Sensecape users had access to these basic diagramming functionalities in addition to the hierarchy view and features (e.g., semantic zoom) described in Section~\ref{section:system}. 

\textit{Design Rationale for Baseline.} Our initial Baseline featured ChatGPT and Miro side by side --- the setup described in Section~\ref{section:motivating_scenario} --- rather than the \textit{integrated} environment described above. We made the change because our pilot study revealed that this was a significantly weaker setup for comparison. Since participants had to transfer all the text from ChatGPT to Miro, whereas Sensecape users did not have to, it was difficult to evaluate the effectiveness of the features introduced in Sensecape. \change{In other words, it introduced substantial app switching costs (e.g., copying text from ChatGPT to Miro), giving an unfair advantage to Sensecape. Moreover, integrated environments are becoming the norm --- e.g., Copilot, Bing Search, Google Docs, Miro, and Figma. Due to these reasons, we developed the Baseline interface, which allowed us to specifically test Sensecape's new features and environment.}


\subsection{Tasks} 
\label{section:tasks}
Participants were asked to use Sensecape and Baseline to explore two topics: (A) \textsc{Impact of AI on the Future of Work} and (B) \textsc{Impact of Global Warming on Economy}. The order of the system and topics was counterbalanced, resulting in 4 (= 2 x 2) conditions, to minimize order bias. They were instructed to imagine they have to give a talk on the topic in two weeks and are using the system to explore the topic and document what they find. To encourage them to organize the collected information, participants were told they would meet with colleagues in the coming week to plan the talk and share the canvas and hierarchy they created in Sensecape. (Full instructions are provided in the Supplementary Material.) 

\subsection{Procedure} After completing the consent form, participants answered demographic questions in a pre-\textit{study} survey. Then they engaged in two tasks. Depending on the condition assigned to them for each task, they used Sensecape or Baseline to explore one of the aforementioned topics. Each task required participants to complete a pre-task survey, a pre-task exercise, a task interface tutorial, the main task, and a post-task survey. The pre-task exercise had two purposes: (1) to assess their prior knowledge of the topic and (2) to acquaint them with basic diagramming functionalities. Participants were instructed to use Sensecape's canvas view, which had only basic diagramming functionalities such as adding, editing, grouping, and connecting nodes with edges, to list any related topics they know or questions they are interested in exploring. Following this exercise, they viewed the system (Sensecape or Baseline) tutorial and undertook practice tasks to familiarize themselves with the assigned system. Then they proceeded with the main task (Section~\ref{section:tasks}) using the assigned system for 20 minutes and then answered a post-task survey to assess its usefulness. After finishing both tasks, they filled out the post-\textit{study} survey, where they indicated their preferred system and rated the usefulness of features in Sensecape. Finally, they participated in an interview to elaborate on their experiences and reasons for their responses. Throughout the study, participants were asked to think aloud. The study was screen recorded for accurate transcription and analysis of their exploration and sensemaking processes. They received a \$30 gift card for this 1.5-hour study.


\subsection{Participants} We recruited 12 participants (age: M = 26.9, SD = 4; gender: 4F, 7M, 1 Prefer Not to Say) from a local R1 university and via mailing list. They had various backgrounds, including computer science, industrial design, neuroscience, and engineering.
Most participants (9 A Lot, 3 Some) had much experience researching complex topics (`searching and making sense of lots of information'). They varied in their experience using online whiteboarding tools (3 A Lot, 6 Some, 3 None) and interacting with / prompting generative AI models such as ChatGPT, new Bing, and DALL-E (4 A Lot, 6 Some, 2 None). Most participants (2 A Lot, 9 Some, 1 None) had experience drawing concept maps, mind maps, or knowledge diagrams. 

\subsection{\change{Measures}}
\change{To observe and analyze the differences in exploration and sensemaking behavior, and perceived utility of suggestions, between Sensecape and Baseline systems, we used the following measures.}

\subsubsection{Exploration Measures}
\change{The search-as-learning and information retrieval communities have consistently used the number of domain-specific terms and the number of nodes in a mind map or knowledge structure as measures for information exploration~\cite{urgo2022learning, zhang2020users}. Following these practices, we used three measures for exploration: (1) \textit{number of prompts} issued; (2) \textit{number of nodes} created in the knowledge structure; (3) \textit{number of concepts} --- the number of unique, relevant, domain-specific concepts on the canvas.}

\change{The number of concepts explored is measured as `the number of unique domain-specific terms based on domain-specific glossaries (e.g., Glossary of AI by Wikipedia).' Two raters coded the number of concepts explored and gathered and had an inter-rater reliability of 0.93 Intraclass Correlation Coefficient (2,1). We did not count the switching between canvases as re-visits to information.}


\subsubsection{Sensemaking Measures} 

\change{Prior literature \cite{urgo2022learning} and the Sensemaking Model by Pirolli and Card \cite{pirolli2005sensemaking} agree that organizing information into schema is an essential step in the sensemaking process. From a cognitive process perspective, this process of structuring information into knowledge hierarchy requires comparing, contrasting, and differentiating new and existing information --- all of which involves revisiting information previously interacted with. Consequently, we used two measures to assess sensemaking. The first is the \textit{number of hierarchical levels} in the knowledge structures. In the Sensecape condition, this included levels of hierarchy on both the canvas and hierarchy views, which allowed us to understand not only the number of concepts explored, but also how participants conceptualized the relationships between concepts. In the Baseline condition, this was the number of levels in their concept maps. The second measure was the \textit{number of revisits} to previous topics. In the Baseline condition, this included scrolling up to read previous chats or clicking and editing parts of the concept map that they had created with previous interactions. In the Sensecape condition, the same interactions were considered.}


\subsubsection{Perceived Utility Measures}
\change{To understand the perceived utility of Sensecape's features, we used responses to the post-study survey and interview as measures. For example, this included responses to questions asking for their agreement (1: Strongly Disagree; 5: Strongly Agree) with statements, such as `The hierarchy view is useful for making sense of complex information.'}
\section{Results}
\label{section:results}
In this section, we report the findings from our analysis of participants' survey responses, think-aloud, and system usage logs. We examined them to understand how Sensecape supports exploration and sensemaking and how participants perceive the utility of its features and the value of Sensecape for their knowledge work.

\begin{figure}[htb!]
	\centering
	\includegraphics[width=0.45\textwidth]{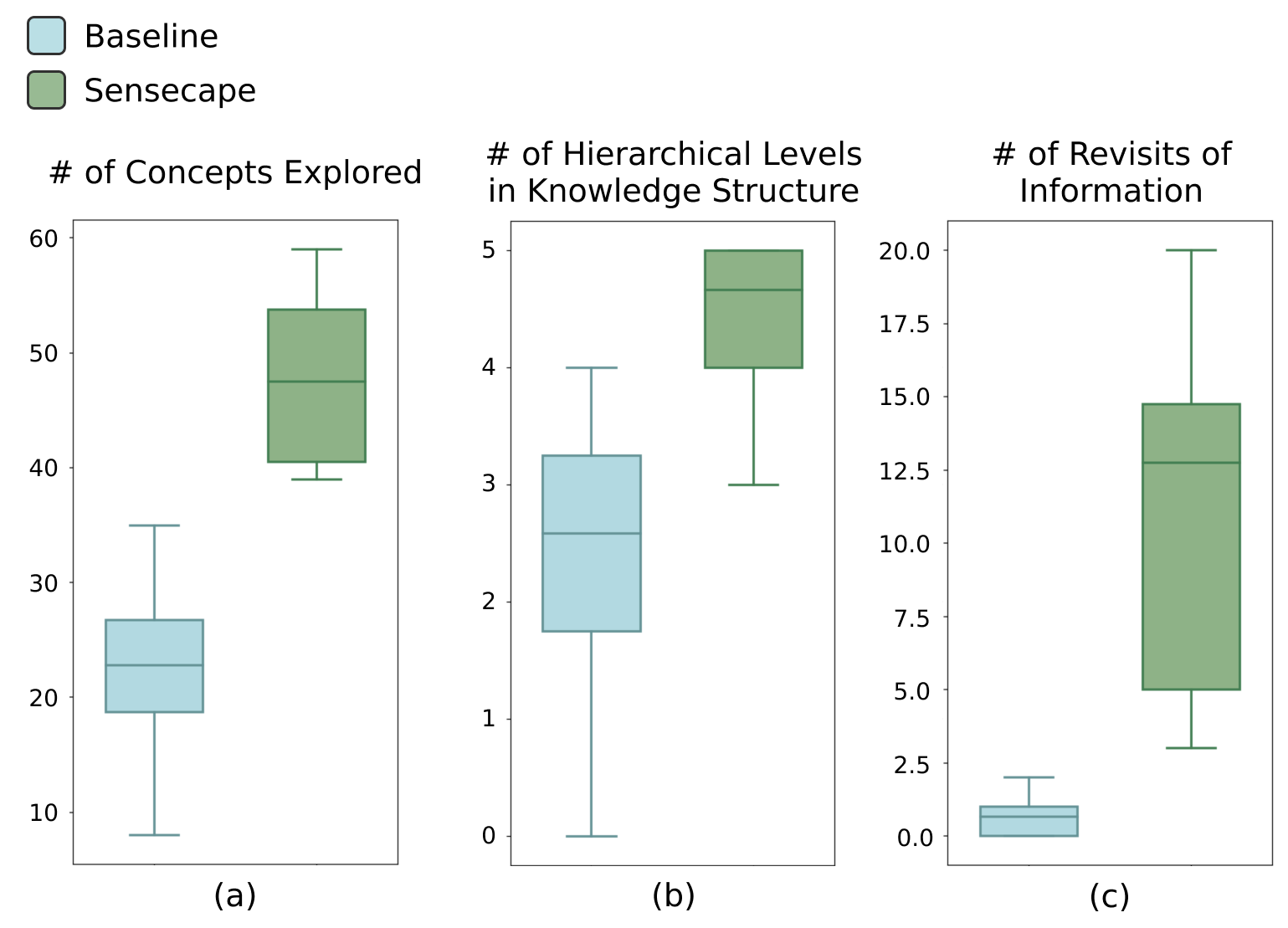}
	\caption{When using Sensecape, participants (a) explored more concepts, (b) structured their knowledge representations more hierarchically, and (c) revisited information they had previously interacted with more frequently.}
	\Description[short description]{long description.}
	\label{fig:log}
\end{figure}

\subsection{RQ1. \change{How does} Sensecape support exploration?}

The analysis of system usage logs shows that participants issued a similar number of prompts in both the Baseline (M = 5.8, SD = 2) and Sensecape conditions (M = 7.3, SD = 5.2), $t(11) = 0.9, p = 0.37$. They also created a similar number of nodes in their knowledge structure: Baseline (M = 23.5, SD = 8.3) and Sensecape (M = 26.8, SD = 10.1), $t(11) = 0.8, p = 0.45$. And made a similar number of connections in their knowledge structure: Baseline (M = 15.8, SD = 9.3) and Sensecape (M = 14.3, SD = 8.5), $t(11) = 0.4, p = 0.72$. 

However, when using Sensecape, participants explored significantly more concepts (M = 68.3, SD = 49.1) than when using the Baseline system (M = 22.8, SD = 7.7), $t(11) = 3.1, p = 0.01^{**}$ (see Fig.~\ref{fig:log}a). 

This suggests that the participants got more information out of issuing a similar number of prompts. P10 described the content of their responses as ``\textit{presenting the littlest amount of information with the most punch}''.  P12 said, ``\textit{I definitely would use [Sensecape] to explore a complex topic because it helps me answer questions, generate content, and helps me automatically lay it out to make sense of}.'' \change{Many participants found expand bar features such as subtopic generation and question generation ``\textit{very helpful}'' in supporting their exploration, as they helped ``\textit{articulate information needs better}'' (P7) and ``\textit{know about concepts or terms [to explore next]}'' (P1). P7 explained: ``\textit{If I were looking on my own, I would not know what to look for. I would Google, I would find an article, then look at multiple articles, and try to pick out common subtopics from there}.''}

\subsection{RQ2. \change{How does} Sensecape support sensemaking?}
\label{section:results-rq2}
When using Sensecape, participants structured their topic knowledge more hierarchically (M = 4.3, SD = 1.2) by adding more hierarchical levels to their knowledge representations than when using the Baseline system (M = 2.6, SD = 1.6), $t(11) = 2.7, p = 0.02^*$ (see \ref{fig:log}b). P9 stated, ``\textit{you can focus your attention on one specific subtopic and dive deeper into each subtopic in a natural way}'', showing that Sensecape provided the participant with a more conducive environment to structure their thinking.  P5 said, ``\textit{it helps me identify connections between topics and reflect on them. This is really helpful to see my knowledge at different levels}.''

Participants also revisited information they had previously interacted with (either as a prompt or in their knowledge representation) more when using Sensecape (M= 12.8, SD = 10.9), compared to when using Baseline (M = 0.7, SD = 1), $t(11) = 3.8, p = 0.00^{**}$ (see \ref{fig:log}c). P11 stated, ``\textit{I actually have that hierarchy in my mind, and it’s actually better to remind me what I have done}.'' P3 talked about how Sensecape helped her ``\textit{easily move from one level to another and better follow [her] structure of thinking and come back to it, to reflect on}.''

\change{The participants also felt that Sensecape helped structure their thoughts. P8 liked the way it ``\textit{helped [him] structure [his] thinking}'' and ``\textit{everything is automatic}'' and he does not ``\textit{need to manually organize the information}.'' P7 also highlighted this by comparing Sensecape with commercial softwares Miro and Google doc:}

\begin{displayquote}
\change{``\textit{I think it helps a lot that each subtopic has its own canvas, but when you are in the hierarchy view, you can still see the subtopics listed within that canvas. This is nice because when it comes to a Miro board, it can get very confusing pretty quickly, so it's nice that each level of information has its own page in Sensecape. On Google Doc, if all the information is set just linearly from top to bottom, it can be hard to find information in the middle. Here you get a sense of where ideas are connected in relation to one another because you're able to lay out things spatially rather than being forced to put things in a linear top to bottom order}.''}
\end{displayquote}

\subsection{RQ3. What is the perceived utility of Sensecape's features?}

\begin{figure*}[htb!]
	\centering
	\includegraphics[width=\textwidth]{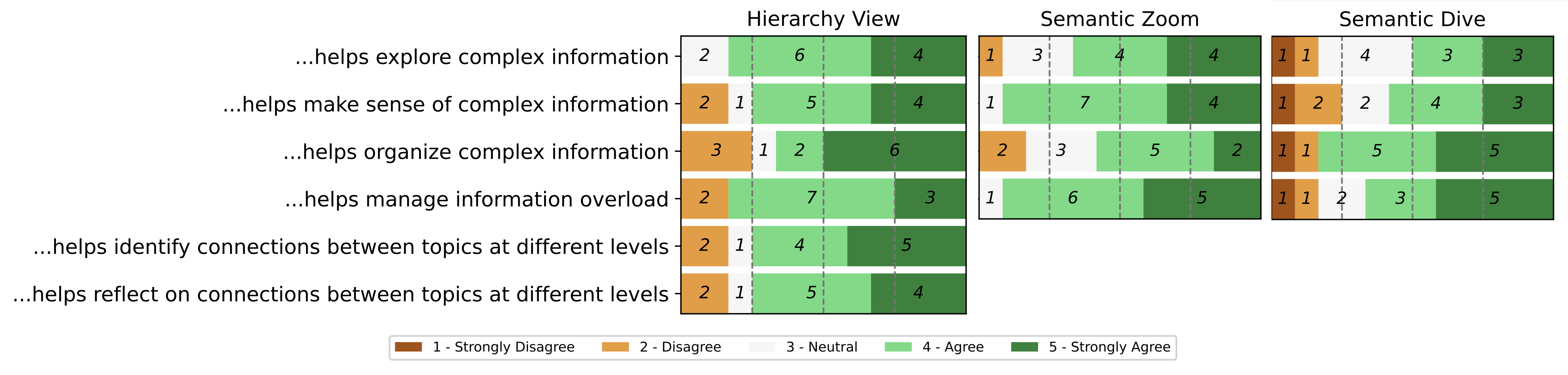}
	\caption{Evaluation of Sensecape's features}
	\Description[short description]{long description.}
	\label{fig:feature-evaluation}
\end{figure*}

We analyzed the responses from the post-study survey and interview to assess the perceived utility of Sensecape's features. 

\subsubsection{Expand Bar} The expand bar allowed users to seek information using \textsc{Prompt}, \textsc{Explain}, \textsc{Questions}, or \textsc{Subtopics} prompts. On average, participants used the expand bar 7.25 times (SD = 5.15) during the study. They used \textsc{Prompt} 4 times (SD = 4.26), \textsc{Explain} 3.41 times (SD = 4.52), \textsc{Questions} 0.41 times (SD = 0.52), and \textsc{Subtopics} 3.83 times (SD = 5.27). 

P7 used the \textsc{Explain} prompt to ``\textit{dive deeper}.'' He said: ``\textit{when you generate a subtopic, you can simply click} \textsc{Explain} \textit{versus writing the prompt, to say, oh }`{\small\fontfamily{lmss}\selectfont jobs rotation}' \textit{maybe it is something I should look more into}.'' P3 found the \textsc{Questions} prompt gave an overview of the topic, saying: ``\textit{I like that it generates questions and gives me an overview on the topic... I struggle to get a big picture of the topic, but with the generated questions I can dive into many different aspects}.'' P7 noted that the \textsc{Subtopics} prompt taught him new ways of exploring and articulating his information needs: ``\textit{without the prompt for generating subtopics, I might not have thought to do that. But now that I’ve been exposed to [Sensecape], I think I would know to ask this to generate more information or more ways of articulating information needs}.'' Similarly, P3 said the \textsc{Subtopics} prompt is ``\textit{great to have because it allows you to get that breadth faster}.''

\subsubsection{Text Extraction} 
To help break down the generated responses, participants chose to extract and curate parts of the response, on average, 6.25 times per session (SD = 5.65). Most participants used this feature to organize the information and their thinking. P6 said, ``\textit{system helped me structure my notes and thinking because of the ability to drag and drop and then organize things spatially on a 2D plane}.'' They further explored these extracted parts of the response, on average, 4.2 times per session (SD = 3.97). This ability to follow up helped P6 dive deeper to explore more. P6 said it ``\textit{helped [her] dive deeper into the subtopics within the topic and allowed [her] to continue asking follow-up questions}.''

\subsubsection{Semantic Zoom} 
On average, participants used semantic zoom 12.58 times per session (SD = 8.92) to manage the information overload from generated responses. When asked to rate the effectiveness of this feature on Likert-type statements, almost all participants agreed that it helped them explore complex information, make sense of, organize, and manage complex information, as shown in Figure \ref{fig:feature-evaluation}. P3 explained her rating: ``\textit{it does allow you to manage information overload because it offers different levels of granularity. You can identify keywords, but also structure information in the summary or in the lines. By traveling across the two, you can get an idea of the complex information}.'' Similarly, P11 said, ``\textit{it did not really directly help me organize, but it gave me some of the views and ideas on how to organize}.'' 

\subsubsection{Semantic Dive} 
On average, participants used the semantic dive feature 6.58 times per session (SD = 5.12). Most participants agreed that the semantic dive feature helped them explore complex information, make sense of, organize, and manage complex information (see Figure \ref{fig:feature-evaluation}). P7 found this feature helped them navigate easily: ``\textit{I think it helps a lot with organizing complex information since you can flow directly between subtopics... versus having to go back out to the hierarchy and back into the topic}.'' P9 identified a trade-off, saying that``\textit{you can focus your attention on one specific subtopic and dive deeper into each subtopic in a pretty natural way. But that might also cost your cognitive load and attention}.''

\subsubsection{Hierarchy View} 
\change{Ten out of 12 participants actively used the hierarchy view to organize information and switch between different levels of abstraction, visiting} it an average of 6.33 times per session (SD = 5.26). Participants stated that the hierarchy view aided in exploring complex information, making sense of it, and managing it. It helped them ``\textit{identify and reflect on connections between topics at different levels}'' (P11). Most participants strongly agreed that the hierarchy view was beneficial in managing information overload (see Figure \ref{fig:feature-evaluation}). \change{P9 highlighted this by comparing Sensecape with an interface without a hierarchy view, saying:} 

\begin{displayquote}
\change{``\textit{The hierarchy view helps distribute the information. If you put all the information on one board, you have to do a lot of grouping, like grouping the concepts together and putting them on different areas of the board. But with the hierarchy view, because it helps you structure the different groups of concepts, you can focus on one concept or idea on one specific board. This allows you to not be distracted by other concepts on the board. You can focus your attention on one specific topic and dive deeper into each node}.''}
\end{displayquote}

Additionally, as mentioned in Section~\ref{section:results-rq2}, participants structured their thinking and topic knowledge more hierarchically in Sensecape using this view. P6 explained, ``\textit{I really like the ability to zoom out of certain canvas. I feel that it helped me organize an even more complex topic than what I have already seen before in other mind mapping tools or relational graphics. The ability to see how each canvas is zoomed out and related with that 3D view was refreshing to me}.'' \change{P4 added: ``\textit{I think people understand these things in a hierarchical way. Definitely, in my mind, when I'm explaining it to someone, I would explain it and understand it hierarchically}.''} \change{Other benefits of the hierarchy view included making the explored information easier to ``\textit{remember}'' (P11) and motivating exploration into unvisited information spaces. P5 mentioned that seeing the hierarchy view taught him ``\textit{how to search, what to search}'' and ``\textit{motivate[d] [him] to search more and explore more complicated information}.'' However, the hierarchy view was perceived as complex and overwhelming by few participants. P3 stated: ``\textit{I think we are not meant to think in three dimensions. Or at least, I am not meant to}.''}


\subsection{RQ4. How do people see Sensecape being useful in their everyday knowledge work?}

\begin{figure*}[htb!]
	\centering
	\includegraphics[trim=0.2cm 0cm 0cm 0cm, clip=true,width=\textwidth]{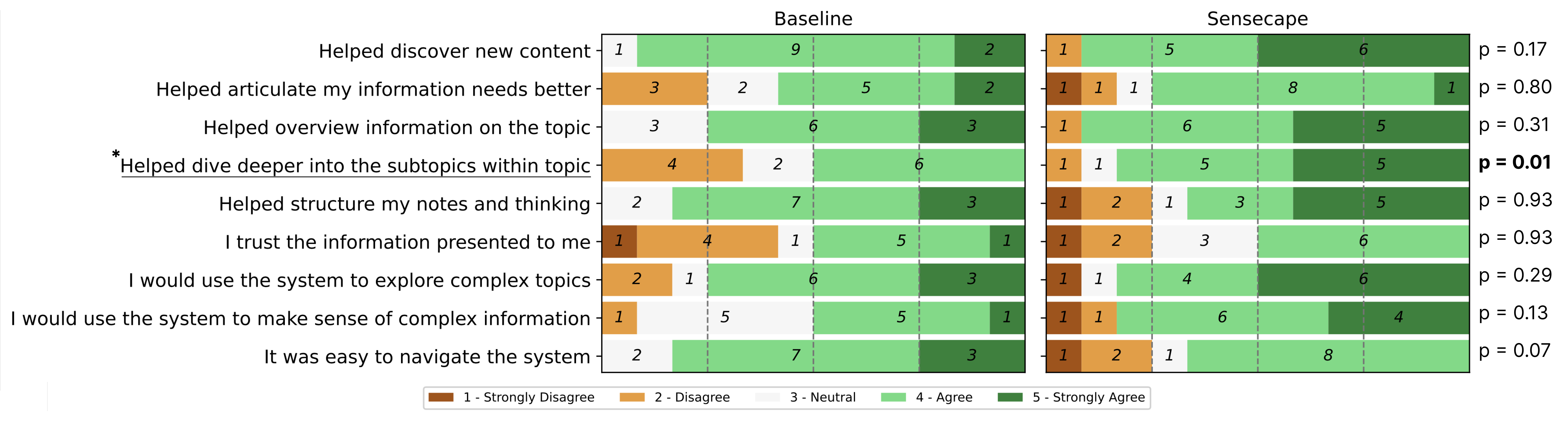}
	\caption{System evaluation results. Statistical significance (p < 0.05) is marked with *.}
	\Description[short description]{long description.}
	\label{fig:system-comparison}
\end{figure*}

To understand how Sensecape might be useful in complex information tasks beyond the scope of our evaluation study, we asked participants \change{---} from diverse fields of knowledge work \change{---} whether they envision Sensecape being beneficial for their work and if so, in what ways. Four of the participants identified as Researchers, four as Designers, and four as Engineers. An analysis of post-study interviews revealed that participants were keen on using Sensecape in various ways for their knowledge work. We discuss these applications in this section.

\subsubsection{To explore and learn about new topics}
Most participants talked about how Sensecape's features could help them explore and learn about new, complex topics in their everyday work. For example, Machine Learning Engineer P8 said, ``\textit{as an engineer, I sometimes need to do something that I didn't know how to, I need to learn some new fields or a new technique. And this is a very good tool for me to explore a new concept. Especially with the hierarchical view, I can navigate the topic much better, and it saves me a lot of time}.''

\subsubsection{To generate novel ideas and develop them}
\label{finding:to-generate-ideas}
Some participants talked about how Sensecape could help them generate ideas and develop them. Researcher P3 spoke about his experience starting a new project in a new domain and how Sensecape might help find relationships between two seemingly disconnected topics: ``\textit{it was challenging for me at first to find a specific topic [for my research]. I knew I wanted to work on invisible illness or chronic illness or disabilities, and I had to tie it in with a piece of pop culture. But I don't watch so much pop culture, and then I didn't know how to tie that into invisible illness. But I could have used this and it could have helped generate ideas. And then, after getting the broad topic, being able to organize it according to subtopic would've been nice. Because in one giant linear block, it was really hard to reason with it or see how the argument is laid out. So I think the system with the hierarchy view will be very helpful with that. It would've also helped me find keywords to look into}.'' Researcher P1 echoed how this might also be useful in a collaborative setting: ``\textit{It's easier to generate ideas, even with collaborators, because it's easy for people to point and say, `I don't think it makes sense', `this concept is related to this picture', or `maybe we should add this in this layer'}.'' 

\subsubsection{To collaborate: share understanding}
\label{finding:to-share-understanding}
Modern knowledge work is often collaborative. While collaboration offers benefits, effectively coordinating work within a team can be challenging. Collaborators need to invest time in dividing and assigning search goals and tasks, locating, sharing, and synthesizing information to create a shared mental model~\cite{shah2010exploring}. Challenges might arise from duplicated efforts among collaborators, as well as confusion about the process and the resultant understanding~\cite{shah2010exploring, capra2010tools}. 

Researcher P1 wanted to use Sensecape to share understanding and their exploration process with their research collaborators: ``\textit{In our research meetings, many times we have to share what we have explored so far in our project. The way we've been doing it is mostly using a document where the information is linearly presented. Having [Sensecape's] visual structure to the knowledge, it'll be easier to guide the discussion. It might also be easier for the other group members to make sense of the overall topics of our discussion}.''

\subsubsection{To collaborate: share process and hand-off}
\label{finding:to-share-process}
Designer P7 shared \change{that Sensecape's ability to allow users to organize information and explicitly reveal connections across different abstraction layers via the hierarchy view can be invaluable when collaborating on intricate tasks, such as software design. P7 remarked:} ``\textit{the web app I'm building has become very complex. So onboarding new software engineers to the project has become very hard. So say, I'm onboarding new engineers and I want to show how one part of the system is designed and the process we used to build it. Then, I think this would be very helpful for showing the connection to the other parts of the system and diving deeper into the specifics of each part}.''

\subsection{Participants Preferred to Use Sensecape for Deeper Understanding of a Topic}

At the end of the study, participants were asked which system they preferred using --- Baseline or Sensecape. All 12 participants favored either Baseline or Sensecape over linear interfaces that lacked an integrated note-taking area, like ChatGPT. Out of the 12, seven participants generally preferred working with Sensecape. Nine participants preferred to use Sensecape when aiming to gain a deeper understanding of a topic. For a broad understanding of a topic, six participants preferred the Baseline, one had no preference, and five preferred Sensecape.

In a post-study survey about participants’ opinions about the two systems, they believed that Sensecape helped them significantly more than the Baseline to `dive deeper into subtopics within a topic'. Participants reported that the Baseline was marginally easier `to navigate.' (The p-values from Wilcoxon signed-rank tests for these statements are reported in Fig.~\ref{fig:system-comparison}.)  
 
Overall, the analysis of interviews revealed that each system's suitability might vary based on the context and the participant, and having the option to switch between the two could be advantageous. P10 remarked, ``\textit{[For getting a] broad overview, [Baseline] would help more, while [for] deeper understanding, [Sensecape] would help more}.'' They believed that with Sensecape, they could delve deeper into suggested keywords and would be equipped to ``\textit{work in more complex environments}.'' However, many noted a trade-off between the learning curve and the complexity of the task at hand. Participants who favored Baseline primarily cited the steep learning curve of Sensecape as the reason. P4 observed, ``\textit{The more advanced features can get more complex things done and much faster, but getting used to them can take some time}.'' Similarly, P12 stated, ``\textit{the learning curve [for Sensecape] is a bit high compared to [Baseline]. So I probably prefer [Baseline] but I can imagine that if there's a huge information that need to be digested, [Sensecape] would be really useful}.'' 
\section{Discussion}
\label{section:discussion}

\subsection{Summary}
In this work, we explored how we can leverage LLMs to support complex information work by developing Sensecape, an interactive system designed to support information exploration and sensemaking in a structured manner. The user evaluation study found that Sensecape enables users to discover more topics and explore more broadly than the Baseline, owing to its numerous features that support exploration through the instant generation of subtopics, questions, explanations, and prompts. Sensecape's expand bar accommodates the wide range of prompts and responses users require for productive information exploration. For instance, when users perform semantic dive, Sensecape instantly recommends subtopics, providing them a head start rather than confronting them with a blank slate, addressing the prevalent challenge faced by individuals with limited knowledge of a subject when searching~\cite{miyake1979ask}.

Furthermore, Sensecape supports participants in developing a deeper understanding of a topic, as evidenced by their self-reports and the multiple levels of hierarchy in their knowledge representation. This may be because Sensecape enables users to externalize their sensemaking in a more nonlinear, hierarchical manner. 
On the other hand, the enhanced sensemaking and exploration seen when using Sensecape might be explained by \textit{schema theory}, which posits that explicitly forming links between new information and the learners' pre-existing knowledge and schemas can facilitate the integration of new information into their schema~\cite{pirolli2005sensemaking, widmayer2004schema}.

\change{Despite its numerous benefits, the interface's complexity also proved challenging for a few participants. For example, while most participants demonstrated the ability to organize and navigate the information space across different levels of abstraction using the hierarchy view within the limited study time frame (\textasciitilde20 min), some found it complex and overwhelming. This observation aligns with prior work that found that 3D visualizations often come with high interface costs and may necessitate time for some users to overcome the initial challenges they encounter~\cite{sebrechts1999visualization}.}


\subsection{Limitations and Future Work}
\subsubsection{Study Limitations.} When we introduce new technologies and tools, the novelty effect can bias people's perceptions of the usefulness of the introduced tools. At the same time, if people have limited time with the tool, they might also not be able to fully assess its usefulness.
For example, P11 noted that a brief interaction with Sensecape might prevent users from recognizing the value of the hierarchy view. P11 suggested that if users are given more time (``\textit{for example, 60 minutes}''), they would ``\textit{get too much information}'' in their canvas and realize the ``\textit{need for the hierarchy view, to organize [their workspace] better}.''
Moreover, it is worth noting that the hierarchy view required participants to think and organize information across multiple levels of abstraction --- a cognitive task that could be unfamiliar and thus challenging for some~\cite{suh2022coding}. 
In fact, participants who preferred Baseline often cited the learning curve of Sensecape as the reason.
While certain participants, like P1, who claim to apply hierarchical thinking in their everyday knowledge work, felt they used Sensecape productively, several participants stated they could not fully grasp and utilize all of Sensecape's features within the provided time.
We leave a longitudinal study spanning several weeks to months in a real-world setting as future work, to test these observations as well as other intriguing research ideas listed below.


\subsubsection{Supporting Switching Between Linear and Nonlinear Interfaces}
\label{sec:switching_between_linear_and_nonlinear_interfaces}
With a greater degree of freedom, users can position objects in various ways in nonlinear interfaces. While this empowers users proficient in navigating them, it can inadvertently challenge those with less experience and skills. This was one of the observations in our study, as a few participants preferred the Baseline interface and found the \textit{option} to leverage additional space (e.g., hierarchy view) rather overwhelming. While linear conversational interaction is also available in Sensecape --- since it appends an input box below the generated response --- users still needed to pan and adjust their view. 
To cater to users with diverse needs and preferences, it might be beneficial to provide an option to switch to Baseline (Fig.~\ref{fig:baseline-interface}) where the conversational interface is \textit{locked in} on the left side of the interface. In fact, P5 alluded to the need for such an option, saying that for a topic he is unfamiliar with, he would like to use Baseline first to collect information, and then when he ``\textit{reaches a particular familiarity, then [he] would move to [Sensecape]}'' and ``\textit{abstract everything.}''

\subsubsection{Enabling Collaborative Multilevel Exploration and Sensemaking}
A significant portion of our information gathering and sensemaking occurs through interactions with others. For example, people pose questions and gather information and opinions from Q\&A platforms like Quora and communities such as Reddit. While the current implementation of Sensecape is not designed to support collaborative exploration and sensemaking, introducing such capabilities could pave the way for innovative methods for people to collectively explore and make sense of the information space through the joint construction of a knowledge hierarchy.
As noted in Sections~\ref{finding:to-share-understanding} and \ref{finding:to-share-process}, our study participants expressed an interest in using Sensecape to collaborate, share the process of conceptual understanding, and perhaps even develop hand-off documentation. 

\subsubsection{Providing Context-Aware Recommendations}
We envision numerous ways to enhance Sensecape, especially with respect to intelligently supporting users during their exploration and sensemaking processes. For instance, we imagine Sensecape analyzing content in users' canvases and hierarchies, and offering content-based recommendations and guidance. Examples of these recommendations may include suggesting potential subtopics or questions when users appear stuck, or even potential groupings and connections between existing nodes. In the hierarchy view, Sensecape could propose potential canvas layers as users browse, or restructure the hierarchy to better align with the granularity of each canvas topic. This proposal could build upon existing research on context-aware recommendations for complex information tasks. Furthermore, since most creative tasks span multiple application contexts, the features and interaction mechanisms of Sensecape could be extended to support intricate information tasks across various applications.

\subsubsection{Using Additional Representations to Represent Information Space}
In Sensecape, we employed a hierarchical representation to encode information. This is because hierarchy is a robust structure for representing knowledge and empowering seamless switching between divergent and convergent thinking inherent in exploratory and sensemaking tasks. While hierarchies are powerful, they are not the only structure suitable for exploration and sensemaking. For example, a graph may more effectively highlight similarities between pieces of information. Thus, in the future, enabling users to leverage more representations could present opportunities to augment and further refine multilevel exploration and sensemaking.

\subsubsection{Enabling Multilevel Exploration and Sensemaking across Diverse Abstraction Ladders}
The spatial exploration in the hierarchy view, particularly the ability to move up and down the levels of abstraction, is inspired by the abstraction ladder~\cite{hayakawa1947language, victor2011, suh2022coding}. 
While Sensecape allows users to transition between the overarching topic and its subtopics, other forms of abstraction ladders do exist. For example, some abstraction ladders employ different representations at each level of abstraction (e.g., code $\Leftrightarrow$ story $\Leftrightarrow$ comic)~\cite{suh2020we, suh2022codetoon}. There are different sets of abstraction levels for each field such as computing (e.g., problem $\Leftrightarrow$ object $\Leftrightarrow$ program $\Leftrightarrow$ execution)~\cite{armoni2013teaching}. It would be interesting to extend this idea of navigating the information space to enable the exploration of different abstraction levels corresponding to different representations and semantic layers used in diverse domains and tasks. Such an extension might aid, e.g., in supporting collaboration in complex tasks like software design and promoting computational \& systems thinking vital for understanding complex systems and processes, and solving complex problems.


\subsubsection{Supporting Fact-Checking.} The LLMs' tendency to ``hallucinate'' has been a large threat to their use in information-related tasks. Recently, researchers and industries have been exploring ways to address this. For example, Microsoft recently launched its search engine, Bing, integrated with ChatGPT. To address the concern of ChatGPT returning incorrect facts, Bing included references to the information sources. Recent start-ups have adopted a similar approach, including references to information sources to mitigate this concern. We believe that such a mechanism should be implemented in all LLM-powered systems that return factual information, and we considered adding it to Sensecape --- such as including links to top search results from the Google search engine. However, since the goal of our study was to assess whether Sensecape facilitates exploration and sensemaking and how it might support other types of complex information work, we intentionally abstained from adding this feature, as it adds an additional distraction during the search process and influences their experience and perception. Nonetheless, for Sensecape to be deployed in real-world settings, there should be mechanisms to verify and determine the extent to which the information generated by LLM can be trusted.





\subsection{Design Implications}
The user challenges identified in this work (Section~\ref{section:motivating_scenario}) and how we address them may have useful implications for the design of future systems that support information tasks with LLMs. Some of these challenges may be specific to exploratory tasks (\textbf{C1. Slow Start}) and conversational interfaces (\textbf{C2. Hard to Revisit}; \textbf{C3. Lack of Structure}). Nevertheless, these insights could still be valuable for instances where one is developing non-conversational but linear interfaces, such as timeline-based interfaces that harness LLMs for information tasks. Conversely, challenges common to a wide variety of systems may be information overload (\textbf{C4. Information Overload}) and visual clutter (\textbf{C5. Visual Clutter}), as the LLMs' ability to instantly generate a large volume of information leaves the user more likely to encounter these issues.
To fully leverage the potential of LLMs, techniques like \textit{semantic zoom} may be necessary. Our study found that participants greatly appreciated features such as semantic zoom and hierarchy view for managing information overload and making sense of complex information.
We envision that a similar type of support, customized to suit specific systems and domains, will be vital for future systems supporting people's complex information tasks with LLMs.
\section{Conclusion}
\label{section:conclusion}

In this work, we introduced Sensecape --- an intelligent interactive system powered by LLMs, designed to facilitate structured information exploration and sensemaking. 
The design of Sensecape is grounded in sensemaking theory and user studies of complex information work.
A user evaluation study found that Sensecape helps users explore more concepts, construct a deeper understanding, and revisit information more frequently to develop a holistic understanding of the complex information space. Besides our contributions of Sensecape and the user study, our work exemplifies how the externalization of multilevel abstraction encourages people to explore further and equips them with powerful tools for exploration and sensemaking of the information space. Collectively, our work offers an exciting initial step towards powering complex information workflows with large language models.

\bibliographystyle{ACM-Reference-Format}
\bibliography{main}

\clearpage
\appendix
\onecolumn

\section{Appendix}
\label{section:Appendix}


\begin{figure*}[htb!]
	\centering	\includegraphics[width=0.88\textwidth]{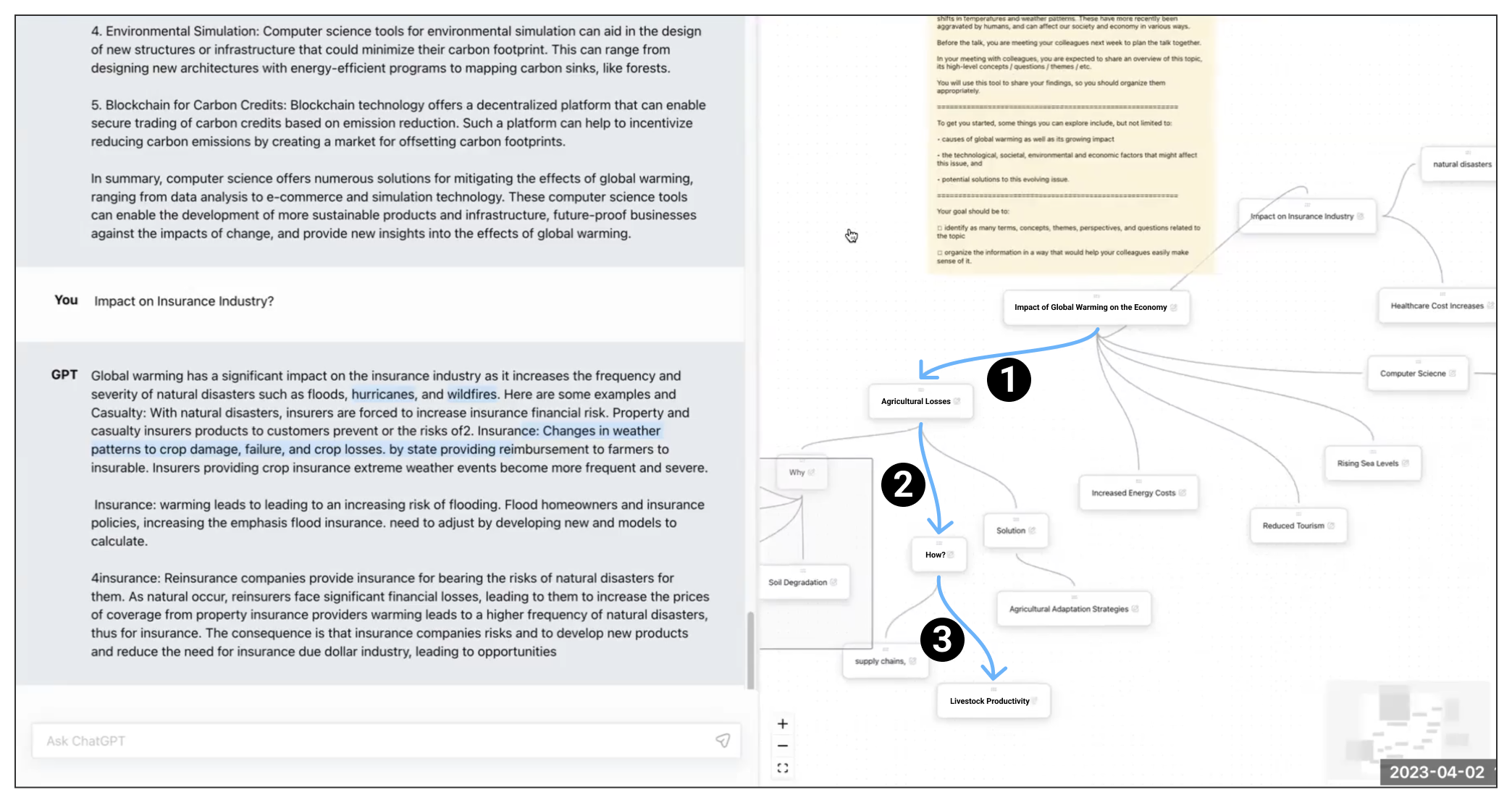}
	\caption{P11's Baseline canvas view: this canvas displays P11's exploration --- to discover the impact of global warming on the economy --- also spreads to agriculture and livestock. P11 traverses through three levels to come to this discovery. The first traversal is shown (1) connecting `{\small\fontfamily{lmss}\selectfont Impact of Global Warming on the Economy}' to `{\small\fontfamily{lmss}\selectfont Agricultural Losses}'. The second traversal (2) connects `{\small\fontfamily{lmss}\selectfont Agricultural Losses}' to `{\small\fontfamily{lmss}\selectfont How?}'. Finally, the third traversal (3) connects `{\small\fontfamily{lmss}\selectfont How?}' to the node `{\small\fontfamily{lmss}\selectfont Livestock Productivity}'.}
	\Description[short description]{long description.}
	\label{fig:p11-baseline}
\end{figure*}

\begin{figure*}[htb!]
	\centering	\includegraphics[width=0.88\textwidth]{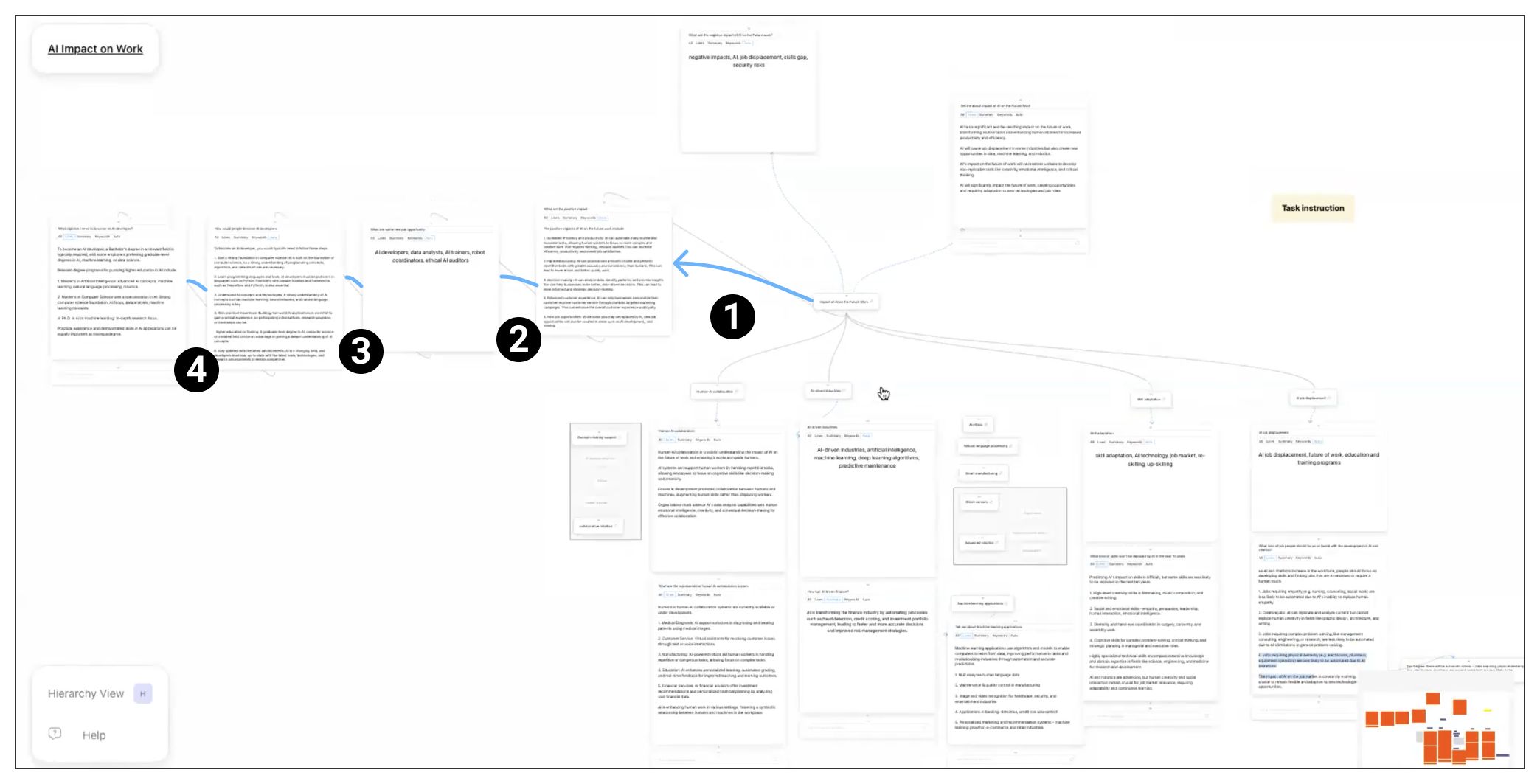}
	\caption{P12's Sensecape canvas view: P12 dove deeper into learning about potential job opportunities in AI. After P12 followed up on the LLM's response three times --- creating four traversals to new ideas and concepts, he took a step back to restructure his canvas. He oriented his most recent conversation towards the left edge of the canvas as indicated by (1) - (4), to position the main topic at the center of all exploration.}
	\Description[short description]{long description.}
	\label{fig:p12-sensecape}
\end{figure*}

\begin{figure*}[htb!]
	\centering	\includegraphics[width=0.85\textwidth]{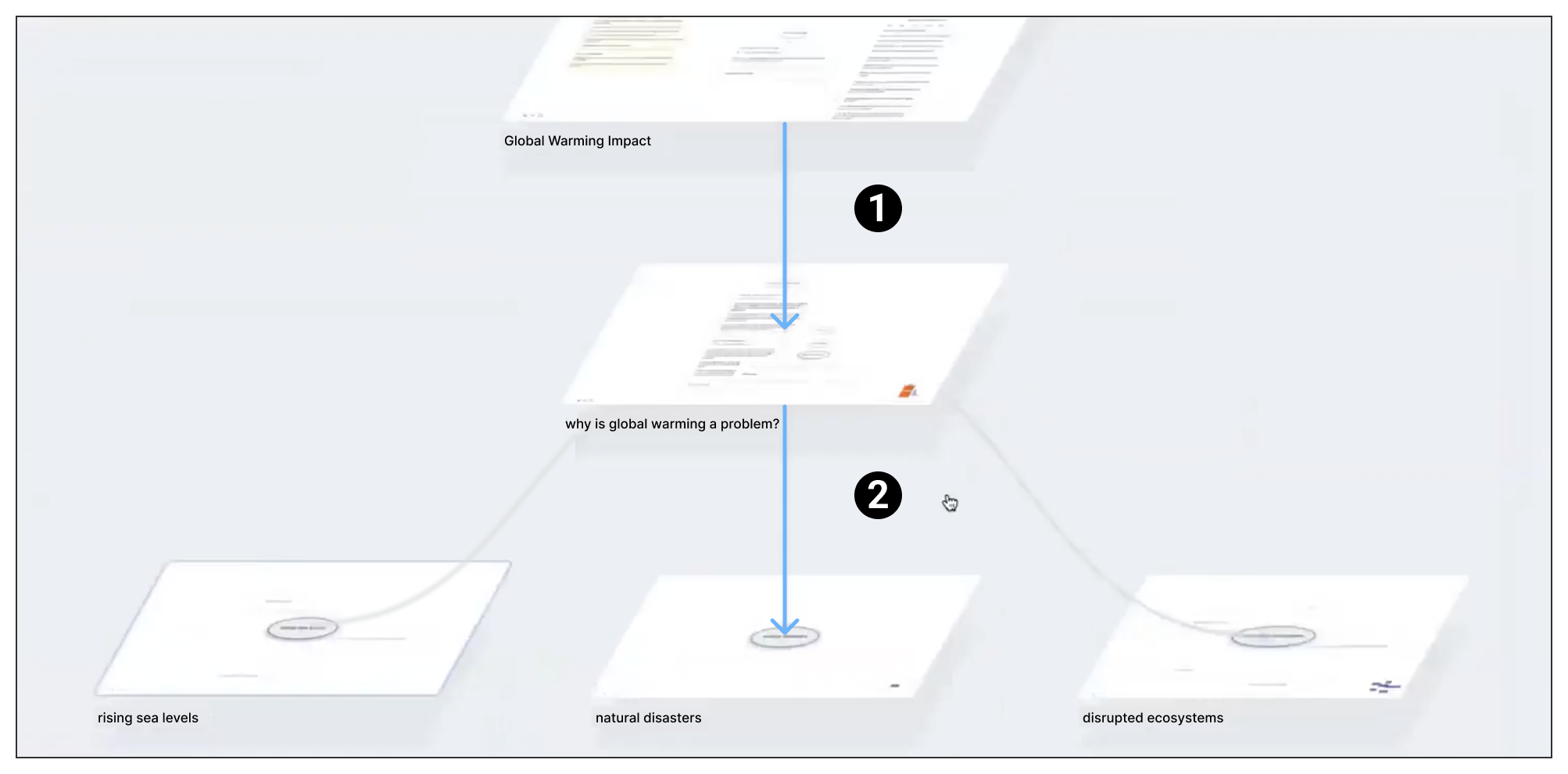}
	\caption{P3's Sensecape hierarchy view: P3 explored the impact of global warming by (1) diving into a question (`{\small\fontfamily{lmss}\selectfont why is global warming a problem?}'). With possible answers organized by subtopics, the participant (2) generated subtopics (`{\small\fontfamily{lmss}\selectfont rising sea levels}', `{\small\fontfamily{lmss}\selectfont natural disasters}', `{\small\fontfamily{lmss}\selectfont disrupted ecosystems}') as answers to the question layer. During the exploration, P3 constructed a hierarchy with two levels to externalize her thought process.}
	\Description[short description]{long description.}
	\label{fig:p3-sensecape}
\end{figure*}

\begin{table*}[h]
    \caption{Prompts used for \textsc{Questions} and \textsc{Subtopics} in Expand Bar and for semantic levels (\textsc{Lines}, \textsc{Summary}, \textsc{Keywords}) in Semantic Zoom. The text with curly braces (e.g., \textcolor{ACMOrange}{\{text\}}) in the `Prompt' column is a placeholder for example input(s).}
    \label{table:llm-prompts}
    \centering
    \resizebox{0.85\textwidth}{!}{
    \begin{tabular}{ p{3.8cm}  p{4.7cm} p{4.7cm}  p{4.4cm} }
        \toprule
\textbf{Prompt Type}      
& \textbf{Prompt}   
& \textbf{Example Input(s)} 
& \textbf{Example Response} \\\midrule
\textsc{Questions}
& \textsf{I need to learn about \textcolor{ACMOrange}{\{text\}}. Give me a total of 25 questions, with 5 questions starting with `why', 5 questions starting with `what', 5 questions starting with `when', 5 questions starting with `where', and 5 questions starting with `how'. Do not add numbers in front of the questions.}
& \textsf{\textcolor{ACMOrange}{Moving to San Francisco}} 
& {\small\fontfamily{lmss}\selectfont Why move to San Francisco?}, {\small\fontfamily{lmss}\selectfont Why is the cost of living so high?}, {\small\fontfamily{lmss}\selectfont Why is San Francisco known as the tech hub?}, ... {\small\fontfamily{lmss}\selectfont What areas offer great value for your money when you are looking for property prices?}  \\\hline
\textsc{Subtopics}        
& \textsf{Give me \textcolor{ACMOrange}{\{numOfTopics\}} give or take \textcolor{ACMOrange}{\{numOfMargin\}} new subtopics in the form of terms in 1 to 3 words each given this context: \textcolor{ACMOrange}{\{context\}}. Format your response in CSV (comma separated values).} 
& \textsf{\textcolor{ACMOrange}{5}, \textcolor{ACMOrange}{0}, \textcolor{ACMOrange}{Fisherman's Wharf}}
& {\small\fontfamily{lmss}\selectfont Pier 39, Street Performers, Seafood Restaurants, Historic Ships, Waterfront Dining.}  \\\hline
\textsc{Semantic Zoom: Lines}
& \textsf{\textcolor{ACMOrange}{\{line\}} If the text stated above is a paragraph, summarize it into a sentence. If the text is a bullet point or numbered list item, keep both the bullet point/number and main topic/term that represented the entire line, but just summarize the description into keywords.}
& \textsf{\textcolor{ACMOrange}{3. Fisherman's Wharf is a popular place to visit for seafood in San Francisco}}
& {\small\fontfamily{lmss}\selectfont Fisherman's Wharf: Fresh seafood, fishermen, Pier 39} \\\hline
\textsc{Semantic Zoom: Summary}
& \textsf{Summarize this text in 1-2 phrases: \textcolor{ACMOrange}{\{text\}}}
& \textsf{\textcolor{ACMOrange}{ Fisherman's Wharf is a popular tourist destination located in San Francisco, California, USA. It is a historic waterfront district that dates back to the mid-1800s, when it was primarily a fishing village.}}
& {\small\fontfamily{lmss}\selectfont Fisherman's Wharf is a popular tourist destination in San Francisco. It was primarily a fishing village.} \\\hline
\textsc{Semantic Zoom: Keywords}       
& \textsf{Extract 3-5 of the most important keywords from this text in CSV format: \textcolor{ACMOrange}{\{text\}}} 
& \textsf{\textcolor{ACMOrange}{Fisherman's Wharf is a popular tourist destination located in San Francisco, California, USA. It is a historic waterfront district that dates back to the mid-1800s, when it was primarily a fishing village.}}
& {\small\fontfamily{lmss}\selectfont Fisherman's Wharf, tourist, San Francisco, fishing village.} \\\hline
        \bottomrule
    \end{tabular}
}
\end{table*}

\end{document}